\documentclass[12pt]{article}
\usepackage{amssymb,amsmath,amsthm,amsfonts,latexsym}
\usepackage{multirow} \usepackage{epsfig} \usepackage{psfrag} 
\usepackage{color}    \usepackage{xcolor} \usepackage{array} 
\usepackage[pdftex,bookmarks, colorlinks, citecolor =blue, breaklinks]{hyperref}
\usepackage{enumitem}

\newtheorem{prop}{Proposition} 
\newtheorem{lema}{Lemma}       
\newtheorem{corol}{Corollary} 
\newtheorem{defin}{Definition} 
\newtheorem{rema}{Remark}

\usepackage{graphicx}
\def\be{\begin{equation}}  \def\ee{\end{equation}}
\def\bea{\begin{eqnarray}} \def\eea{\end{eqnarray}}

\def\Nset{\mathbb{N}}

\def\Sset{\mathbb{S}}
\def\Tset{\mathbb{T}}
\def\Rset{\mathbb{R}}  
\def\Zset{\mathbb{Z}}

\def\Kcal{\mathcal{K}}

\def\Ocal{\mathcal{O}}
\def\Pcal{\mathcal{P}}
\def\Wcal{\mathcal{W}}

   \def\eps{\varepsilon}
      \def\vp{\varphi}
\def\om{\omega}          

  \newcommand{\fin}{\hfill $\Box$}
\newcommand{\fns}{\footnotesize} 

\def\Mic{{M_{\vp,a}}}	
\begin{document}
\begin{center}
\noindent {\Large\bf
Accelerator modes and anomalous diffusion in 3D volume-preserving maps
}
\end{center}

\begin{center}
{\large James D. Meiss$\,^1$, 
        Narc\'{\i}s Miguel$\,^2$, 
        Carles Sim\'o$\,^2$ and 
        Arturo Vieiro$\,^2$}
\vspace*{1mm}

{\footnotesize
\noindent $\,^1$Department of Applied Mathematics, University of Colorado
Boulder, CO 80309-0526, United States\\
\noindent $\,^2$Departament de Matem\`atiques i Inform\`atica, Universitat de
Barcelona, Gran Via 585, 08007, Barcelona, Catalunya}

\tt 	 {jdm@colorado.edu}, 
\tt      {narcis@maia.ub.es}, 
\tt      {carles@maia.ub.es},  
\tt      {vieiro@maia.ub.es}

\end{center}

\begin{center}\today\end{center}

\begin{abstract}
Angle-action maps that are periodic in the action direction
can have accelerator modes: orbits that are periodic when projected onto the torus, but that lift to
unbounded orbits in an action variable. In this paper we construct a 
volume-preserving family of maps, with two angles and one action, 
that have accelerator modes created at Hopf-one (or
saddle-center-Hopf) bifurcations. Near such a bifurcation we show that there is often a
bubble of invariant tori.
Computations of chaotic orbits near such a bubble 
show that the trapping times have an algebraic decay similar to that seen
around stability islands in area-preserving maps. As in
the 2D case, this gives rise to anomalous diffusive properties of the action in our 3D map.
\end{abstract}


\section{Introduction}\label{sect:intro}

In this work we consider real-analytic, volume-preserving maps (VPM) $F$ on the
cylinder $\Tset^d\!  \times \Rset^l$, where $\Tset^d =
\Sset^1\times\!\!\stackrel{d}{\cdots}\!\!  \times\Sset^1$, and $\Sset^1=
\Rset/\Zset$. We think of the variables $(x,z) \in \Tset^d\!\times \Rset^l$ as
being $d$-angles and $l$-actions, and call $F$ an \textit{angle-action} map.
As an important, non-generic property, we will assume that $F$ is the lift of a
smooth map $\tilde{F}$ on the torus $\Tset^d\!\times \Tset^l$; that is, we
assume there is a projection
\begin{equation}\label{eq:projection}
	\Pi:\Tset^d\!\times \Rset^l\to \Tset^{d+l} ,
\end{equation}
such that for each point in $\Tset^d\!\times \Rset^l$,
\begin{equation}\label{eq:lift}
	\tilde F \circ \Pi = \Pi \circ F .
\end{equation}
We will simply take $\Pi(x,z) = (x, z \mod 1)$: the unit modulus is applied to
each action variable.  Such maps may have special orbits, called
\textit{accelerator modes} that are unbounded orbits of $F$ whose projections
onto the torus become periodic orbits of $\tilde{F}$  \cite{Chi79,CI73,Kar82,
Zas97, RKZ99}.  The interest in such orbits is due to the fact that they can
have a huge impact on the properties of chaotic orbits that are unbounded in
the action direction---namely normal diffusion can become super, or anomalous,
diffusion \cite{LL92}.  The way these diffusive properties change due to the
presence of accelerator modes depends on the local structure of the phase space
near the projected periodic orbit. And, as we will see, for finite-time
simulations the statistics outside this local structure also plays a leading
role. 

Throughout this paper we label an orbit of $F$ by subscripts, so that $(x_{t+1},z_{t+1}) = F(x_t,z_t)$.

Accelerator modes have predominantly been studied for area-preserving
maps\footnote
{But some higher-dimensional symplectic maps have also been studied, see
\cite{KM90}.}
defined on $\Sset^1\times\Rset$ ($d=l=1$ above), as exemplified by Chirikov's
standard map  \cite{Chi79}
\begin{eqnarray}\label{eq:stdmap}
C_k:\Sset^1\times\Rset\to\Sset^1\times\Rset,
&\quad&
C_k:
\left(\begin{array}{c}x\\z\end{array}\right)
\mapsto
\left(\begin{array}{c}x'\\z'\end{array}\right)
=
\left(\begin{array}{c}x+z'\\z+k\sin(2\pi x)\end{array}\right).
\end{eqnarray}
As Chirikov showed, when the parameter $k=n\in\Nset_+:=\Nset\setminus\{0\}$ there are $2n$
accelerator-mode orbits 
$$
C_n(\tfrac14,p)=(\tfrac14,p+n),\quad C_n(\tfrac34,q) = (\tfrac34,q-n),\qquad
p,q\in\Zset,
$$ 
that project onto two fixed points of $\tilde{C}_n$ located at
$P_1=(\tfrac14,0)$ and $P_2=(\tfrac34,0)$. When $k$ is an integer these fixed
points are parabolic. This bifurcation can be unfolded, as it will be
explained in \S\ref{sect:apm}. When $\kappa_n=k-n >0$, is small,
there appear islands of stability. Chaotic orbits outside these islands of
stability may be trapped nearby for many iterations, a phenomena known as
\textit{stickiness} \cite{Kar82, Kar83}.  In the map $C_k$ this produces large
excursions in the action variable $z$.\\

The aim of this paper is to generalize this phenomenon to three-dimensional,
volume-preserving maps (VPM) with two-angles and one-action. More concretely, we will:
\begin{enumerate}
	\item {\sl Construct a one-parameter family of VPM of the cylinder
$\Tset^2\times\Rset$ that has accelerator modes} (see \S\ref{sect:model3D}).

We restrict ourselves to a family $F_\eps
:\Tset^2\times\Rset\to\Tset^2\times\Rset$ of the form:
\begin{eqnarray}\label{eq:tipusmap}
F_\eps:
\left(\begin{array}{c}x\\y\\z\end{array}\right)\mapsto
\left(\begin{array}{c}x'\\y'\\z'\end{array}\right)=
\left(\begin{array}{c}x+\Omega_1(z)\\y+\Omega_2(z)\\z\end{array}\right)+
\eps \left(\begin{array}{c}h_1(x,y,z)\\h_2(x,y,z)\\h_3(x,y,z)\end{array}\right).
\end{eqnarray}
The preservation of volume is imposed as $\det DF_\eps(x,y,z) \equiv 1$.  The
generalization from 2D to 3D will be done by constructing the family to mimic
some features of Chirikov's map \eqref{eq:stdmap}. Namely:
\begin{enumerate}
	\item The parameter $\eps$ in \eqref{eq:tipusmap} represents the
deviation from integrability. For $\eps = 0$ all orbits lie on 2D rotational
invariant tori (RIT), $\{(x,y,z): z=z_0\}$, and the dynamics is simply a rigid
rotation in the angles with rotation vector $\Omega(z_0) = (\Omega_1(z_0),
\Omega_2(z_0))^\top$. For $\eps > 0$, but small, $F_\eps$ is assumed to satisfy
the hypotheses of the KAM-like theorems for volume-preserving maps
\cite{CS89,Xia92}. Hence there is a Cantor set of RIT. 
	\item Accelerator modes of $F_\eps$ are born at $\eps=n\in\Nset_+$.
These project to isolated fixed points of $\tilde{F}_\eps$.
	\item When  $0<\eps-n\ll1$ there is a neighborhood of some of the accelerator modes that contains
a bubble of trapped orbits that exhibit regular motion.
\end{enumerate}
The requirement (c) is mandatory since we are interested in accelerator modes
that give rise to anomalous diffusion along the $z$ coordinate. To ensure this,
we will assume that the parameter $\kappa_n=\eps-n$ unfolds a ``Hopf-one" or
``saddle-center-Hopf" bifurcation at the accelerator modes.  This bifurcation,
a discrete analogue of the Hopf-zero bifurcation for ODEs, corresponds to the
creation of a fixed point with multipliers $\lambda_1 = 1$ and $\lambda_{2,3} =
e^{\pm 2\pi i \omega}$ on the unit circle.  The unfolding of this bifurcation
gives rise to a pair of saddle-focus fixed points.  There is a rich structure
around the stability region where orbits may be trapped for a long time so
that they affect the diffusion in the action variable. See \S\ref{sect:hopfone}
for more discussion.

The proposed family of maps $F_\eps$ seems to be the first studied example of
VPM with accelerator modes. 

	\item {\sl Study the effect of these accelerator modes on the diffusive
properties of the action} (see \S\ref{sect:diffusion3D}).

We perform a numerical exploration based on long-term simulations of $F_\eps$
to study, on the one hand, the diffusive properties of the action, and on the
other hand, the trapping statistics due to the passages near the stability
region that appears in a vicinity of the accelerator modes. Here, by trapping
statistics we mean the distribution of trapping times in a neighborhood of the
accelerator mode stability region. Our experiments suggest that this behaves as
$t^{-b}$, $b\in(2,3)$, which is consistent with the behavior observed in the
area-preserving case. Furthermore, the action exhibits an anomalous,
super-diffusive behavior.

\end{enumerate}

This paper is organized as follows. In \S\ref{sect:prelim} we recall some
preliminary facts and set the problem in the proper context by discussing the
well-known analogous area-preserving setting. We summarize some relevant facts
on the Hopf-one bifurcation in the volume-preserving context. The rest of the
paper is separated into two distinct parts according to the previous
enumeration. In \S\ref{sect:model3D} we construct a family of VPM with
accelerator modes and we study the scaling properties of the local dynamics.
In \S\ref{sect:diffusion3D}, we numerically study the diffusive properties and
trapping statistics due to these accelerator modes for an example.  In
\S\ref{sect:discussion3D} we discuss these results, taking into account
geometrical and statistical facts.  Finally, in \S\ref{sect:conclusion} we
summarize our results and propose new lines of research that emerge from this
study.

\section{Preliminaries}\label{sect:prelim}

In this section we introduce the main ideas on which this paper is based. In
\S\ref{sect:apm} we review well-known facts about the accelerator modes of
Chirikov's standard map: the mechanism of their creation, their local dynamics,
and their effect on the action diffusion due to stickiness. This map serves as
inspiration for the construction of our main model. In \S\ref{sect:defam} we
generalize the concept of accelerator mode to higher-dimensional maps.
In \S\ref{sect:shears3D} we define a VPM that can possess accelerator modes as a composition
of simple shears. We finish this preliminary
section by reviewing some facts on the Hopf-one bifurcation in
volume-preserving maps in \S\ref{sect:hopfone}.  This is a mechanism that can
create a region of stable motion in a vicinity of an accelerator mode.

\subsection{Accelerator modes in area-preserving maps}\label{sect:apm}

One of the most studied area-preserving models with accelerator modes is
Chirikov's standard map $C_k$ \eqref{eq:stdmap} \cite{Chi79}. This map has
three properties that we will generalize to higher dimensions.
\\

\noindent{1. \sl Accelerator modes}. As we noted in
\S\ref{sect:intro}, the backward and forward orbits of the
points $P_1=(\tfrac14,0)$ and $P_2=(\tfrac34,0)$ are unbounded under $C_n$ for $n\in\Nset_+$.  
These points are unstable, parabolic fixed points of the projection
$\tilde{C}_n$, and their properties are equivalent under a reflection symmetry
of the map. The parameter $\kappa_n=k-n$ unfolds a saddle-center bifurcation at
$P_1$ (resp. $P_2$) giving rise to an elliptic fixed point $P_{1,e}$ and a
hyperbolic fixed point $P_{1,h}$ (resp. $P_{2,e}$ and $P_{2,h}$) of
$\tilde{C}_{n+\kappa_n}$.  The positions of these fixed points depend on
$\kappa_n$, but, to simplify the notation, we do not make this explicit.
These four fixed points are projections of accelerator modes of $C_k$.\\

\noindent{2. \sl Stability islands around elliptic accelerator modes and limit
local dynamics.} When $0 < \kappa_n \ll 1$, islands of stability appear around
$P_{1,e}$ and $P_{2,e}$. The area of these islands decreases
with $n$ as $1/n^2+\Ocal(n^{-6})$.
The orbits $P_{1,e}$ and $P_{2,e}$ undergo a period-doubling bifurcation
at $k=n+2/(n\pi^2)+\Ocal(n^{-3})$.  Chirikov and Izraelev
\cite{CI73} showed that these scalings hold for a larger class of maps (where
the force $\sin(2\pi x)$ in \eqref{eq:stdmap} is generalized to an odd periodic
function of $x$). In \cite{MSV13} it was proved that the leading terms of the
suitably scaled Taylor expansions of $C_n$ around the accelerator modes could
be written as the quadratic area- and orientation-preserving H\'enon map (which
we just call the H\'enon map from now on) in Karney's form \cite{Kar82}
\begin{eqnarray}\label{eq:karneymap}
H_\kappa:
\left(\begin{array}{c}\xi\\\eta\end{array}\right)
\mapsto
\left(\begin{array}{c}\xi'\\\eta'\end{array}\right)
&=&
\left(\begin{array}{c}\xi+\eta'\\\eta+\kappa-2\pi^2\xi^2\end{array}\right) .
\end{eqnarray} 
The corrections to this map are $\Ocal(n^{-2})$ in each variable. Hence the
H\'enon map becomes asymptotically accurate as $n\to\infty$.  Furthermore,
the coefficients of $\Ocal(n^{-2})$ corrections are small \cite{MSV15}, so
that even when  $n = 1$, the H\'enon map is a fairly good approximation.\\

\noindent{3. \sl Statistics of chaotic orbits in the presence of accelerator
modes}. The stability islands around the accelerator modes are responsible for the
anomalous transport of the action of $C_{n+\kappa_n}$.
There are two interconnected problems of interest in this situation.
Let us restrict ourselves to the island around $P_{1,e}$, though by
the reflection symmetry, the following also applies to the island
around $P_{2,e}$.
\begin{enumerate}[label=(\alph*)]
	\item {\sl Trapping statistics around stability islands}. Let $\Kcal$ be
a compact subset  of the phase space that contains the stability island around
$P_{1,e}$ for $\tilde{C}_k$. Initial conditions in $\Kcal$ that are
not confined by an invariant curve of the stability island or any of its
satellites will escape from $\Kcal$, but have a trapping probability that decays
asymptotically as $t^{-\gamma}$, where $\gamma\in(1,2)$ \cite{Kar82, Kar83, MeiC83, HCM85,
MO86, Zas97, CK08, Ven09, CH10, MSV15}. Equivalently, the density of the exit-time distribution
$\Pcal_k(t)$, the probability that an orbit leaves $\Kcal$ after exactly $t$
iterates \cite{Mei15},  decays as
\begin{eqnarray}\label{eq:trappingstat}
\Pcal_k(t)\sim t^{-b},\qquad b\in(2,3),
\end{eqnarray}
where $b = 1 + \gamma$ and $\sim$ denotes asymptotic behavior as $t \to
\infty$.  The numerical simulations---for finite times---show that $b$ depends
on $k$. Note that the probability density $\Pcal_k(t)$ has bounded average but
all higher-order moments are unbounded. 
	\item {\sl Anomalous diffusion of the action}. The action diffusion
	is computed from the standard deviation $\sigma_T(k)$ of the action $z$ after $T$ 
iterates over an ensemble of orbits that are not confined in stability islands. 
Without accelerator modes, one expects \cite{LL92}
$$
\sigma_T(k)\sim \sqrt{T},
$$
but when there is an elliptic accelerator mode, for example, when
$\kappa_n\in(0,2/(n\pi^2)+\Ocal(n^{-3}))$, one observes super-diffusion:
\begin{equation}\label{eq:stdevdif}
	\sigma_T(k)\sim T^{\chi}, \quad {\chi}>\tfrac12.
\end{equation}
Again, it is observed that the exponent $\chi$ depends on $k$ in a complicated way. 
\end{enumerate}

The dependence of the exponents $b$ and $\chi$ on $k$---for finite time
simulations---is primarily due to the structure of the invariant sets (Cantori,
satellite islands, etc.) surrounding the main accelerator-mode island \cite{MO86}. 
The variation of the exponents is most prominent just after the
breakdown of an outermost invariant curve that had confined a large region of
chaos. The corresponding values of $\kappa_n$ where larger variations are expected
are related to the breakdown of the invariant
curves around elliptic periodic islands of moderate period, as can be seen in the H\'enon map \cite{MSV13}. 
Even though this geometrical fact is well known, and forms the basis for
most of the models of trapping statistics \cite{MO86, CK08, Ven08, OFM16, OFM17}, it is
still not completely understood theoretically and requires extensive numerical
explorations for confirmation.  We refer to \cite{CS84, CS84b, Zas97, MSV15}
for dedicated numerical explorations focusing on concrete Cantori with a
prescribed rotation number.

It is natural to think that the exponents in $\sigma_T(k)$ and $\Pcal_k(t)$ are
related. Under some simplifying assumptions, it has been shown that $2\chi +b =
4$ \cite{Kar83}. This was also
later derived in \cite{GZR88, IHKM91, ZK93, ZK94, WS96}, see the review
\cite{AK08} and references therein. A similar result, obtained in
\cite{MSV15}, shows that $\sigma_T(k)$ is bounded from below 
by $T^{2-(b+1/b)/2}$ for large enough $T$.

\subsection{Accelerator modes for higher-dimensional maps}\label{sect:defam}

As in \S\ref{sect:intro}, let $F:(x,z)\mapsto(x',z')$ be a volume-preserving
map of $\Tset^d \times \Rset^l$ that smoothly projects to a map $\tilde{F}$ on
the torus $\Tset^d \times \Tset^l$, as defined by \eqref{eq:lift}.

As in the area-preserving case, an accelerator mode of $F$ is an orbit with
unbounded action that projects to a periodic orbit of $\tilde{F}$, due to the
periodicity of the map in the action direction.  This implies that the action
increases linearly under iteration of $F$.

\begin{defin}\label{def:accelerator} 
The orbit of a point $(x, z)$ under $F$ is an {\rm accelerator mode} if there
exist $q\geq1$ and $n\in\Zset^l \setminus \{0\}$ such that $F^q(x,z)=(x,z+n)$.
\end{defin}

Note that the projection of an accelerator mode is a $q$-periodic orbit of
$\tilde{F}$. In \S\ref{sect:shears3D} we present a simple way to generate VPM
on the cylinder $\Tset^2\times\Rset$ with accelerator modes. We are mainly
interested in those accelerator modes that project onto fixed points of
$\tilde{F}$, i.e., for $q = 1$. We refer to these kind of orbits as ``fixed point" accelerator
modes, or simply FPAM.

\subsection{Volume-preserving maps as compositions of
shears}\label{sect:shears3D}

To ease the construction of volume-preserving maps, we will consider
angle-action maps that are compositions of shears. Let $S_i$ be a shear in the
$i^{th}$ direction, that is, if $w=(x,z)\in \Tset^{d+l}$, then $S_i:
\Tset^{d+l} \to \Rset^{d+l}$ is 
$$
	S_i(w) = w + s_i(w)\,\hat{e}_i,
$$
where $s_i:\Tset^{d+l}\to \Rset$ is a smooth function that is independent of
the $i^{th}$ component, $w_i$, and $\hat{e}_i$ is the $i^{th}$ unit vector in
the canonical basis of $\Rset^{d+l}$. Assuming that $s_i$ projects to a smooth
function on the circle $\Rset/\Zset$, then $S_i$ projects to a smooth, volume
and orientation preserving map, $\tilde{S}_i$, on $\Tset^{d+l}$. Thus any
composition $\tilde{F} = \tilde{S}_{i_1} \circ \tilde{S}_{i_2} \circ \ldots
\circ \tilde{S}_{i_j}$ with arbitrary $j\geq1$ and
$i_1,i_2,\ldots,i_j\in\{1,2,\ldots,d+l\}$ is a volume-preserving map on the
$d+l$-torus.

In this paper, we are interested in the dynamics of a volume-preserving map
with two angles $(x_1,x_2) = (x,y)\in\Tset^2$ and one action $z\in\Rset$, and
we will use three shears, one in each direction:
\begin{eqnarray*}\label{eq:shears3D}
&
S_1\!:\!\!
\left(\begin{array}{c}\!\!x\!\!\\\!\!y\!\!\\\!\!z\!\!\end{array}\right)
\!\!\mapsto\!\!
\left(\begin{array}{c}\!\!x+s_1(y,z)\!\!\!\\y\\z\end{array}\right)\!\!,
\,\,S_2\!:\!\!
\left(\begin{array}{c}\!\!x\!\!\\\!\!y\!\!\\\!\!z\!\!\end{array}\right)
\!\!\mapsto\!\!
\left(\begin{array}{c}x\\\!\!y+s_2(x,z)\!\!\\z\end{array}\right)\!\!,
\,\,S_3\!:\!\!
\left(\begin{array}{c}\!\!x\!\!\\\!\!y\!\!\\\!\!z\!\!\end{array}\right)
\!\!\mapsto\!\!
\left(\begin{array}{c}x\\y\\\!\!z+s_3(x,y)\!\!\end{array}\right).
&
\end{eqnarray*}
There are two sets of conjugate maps formed by composition of these three
shears in some order, but the families are equivalent under permutations of the
labels. To fix ideas, we let $\tilde{F}=\tilde{S}_2\circ \tilde{S}_1\circ
\tilde{S}_3$,
\begin{eqnarray}\label{eq:tipus}
\tilde{F}:
\left(\begin{array}{c}x\\y\\z\end{array}\right)\mapsto\
\left(\begin{array}{c}x'\\y'\\z'\end{array}\right)&=&\
\left(\begin{array}{c}x+s_1(y,z')\\y+s_2(x',z')\\z+s_3(x,y)
\end{array}\right) \mod 1.
\end{eqnarray}
We will assume that the functions $s_i$ are either periodic or degree-one
functions of their arguments.  In this case periodic orbits of $\tilde{F}$ on
$\Tset^3$ may not be periodic orbits of $F$, the lift to $\Tset^2 \times
\Rset$: the lifted $z$ variable may increase or decrease by an integer amount
in $q$ iterates for suitable $(x,y)$.  Thus for an FPAM, there must exist
points $(x_0,y_0,z_0)$ such that $F(x_0,y_0,z_0) = (x_0,y_0,z_0+n)$ for some
nonzero integer $n$. 

The inverse of the map \eqref{eq:tipus} is simply given by $\tilde{F}_\eps^{-1}
= S_3^{-1}\circ S_1^{-1}\circ S_2^{-1}$, where the three inverses $S_j^{-1}$,
$j=1,2,3$ are obtained by simply changing the sign of the functions $s_i$.

In \S\ref{sect:choiceshears} we will obtain a one-parameter family of maps $F_\eps$, by
letting $s_3(x,y) \to \eps s_3(x,y)$. The existence of an FPAM will then depend upon the parameter $\eps$.

\subsection{The Hopf-one bifurcation in volume-preserving
maps}\label{sect:hopfone}

Suppose that the map \eqref{eq:tipus} has an accelerator mode. In this section
we will add extra conditions on $F$ to ensure that the corresponding periodic
orbit of $\tilde{F}$ has a neighborhood of stable motion. For area-preserving maps,
stable motion around accelerator modes is generated by
a saddle-center bifurcation, recall \S\ref{sect:apm}. A generalization of
this mechanism  to VPM is the codimension-two, \textit{Hopf-one} or
saddle-center-Hopf bifurcation \cite{DM08, DM09}. This bifurcation is the
discrete-time, volume-preserving version of the Hopf-zero, fold-Hopf or
Gavrilov-Guckenheimer bifurcation \cite{GK07}.

To guarantee that there is a stability region near an accelerator mode that is
born from a Hopf-one bifurcation, we will require that the leading terms of the
Taylor expansion give a map that is locally conjugate, using a suitable
scaling, to a map, $\Mic: \Rset^3 \to \Rset^3$, of the form
\begin{eqnarray}\label{eq:Michelson_map}
\Mic:
\left(\begin{array}{c}
u\\v\\w
\end{array}\right)
\mapsto
\left(\begin{array}{c}
u'\\v'\\w'
\end{array}\right)
=
\left(\begin{array}{c}
u+\vp v\\
v+\vp w'\\
w+\vp\left(1-u^2-av\right)
\end{array}\right),
\end{eqnarray}
for suitable values of the parameters $\vp$ and $a$. This map can be regarded
as a 3D analogue of the H\'enon map \eqref{eq:karneymap} since {(a)} it is a
quadratic truncation of the unfolding of the normal form near a triple-one
multiplier \cite{DM09}, {(b)} its inverse is also a quadratic volume-preserving
map~\cite{LM98}, and {(c)} it appears as a truncation of the return map near a
homoclinic quadratic tangency \cite{GMO06}.

The map \eqref{eq:Michelson_map} is a discretization of the well-known
Michelson ODEs \cite{Mic86}
\begin{eqnarray}\label{eq:Michelson_flow}
	\frac{du}{dt} = v,\quad 
	\frac{dv}{dt} = w, \quad 
	\frac{dw}{dt} = 1 - u^2 - av,\qquad a>0,
\end{eqnarray}
that appear in travelling wave solutions of the Kuramoto-Sivashinsky PDE.  The
flow of \eqref{eq:Michelson_flow} has an ``integrable'' limit for $a\to\infty$.
To see this, it is convenient to introduce the scaling $u =\xi$, $v =
\sqrt{a}\eta$, $w = a \zeta$, $t = \tau/\sqrt{a}$. Then
\eqref{eq:Michelson_flow} reads
\begin{eqnarray}\label{eq:Michflowsca}
	\frac{d\xi}{d\tau} = \eta,\quad 
	\frac{d\eta}{d\tau} = \zeta,\quad 
	\frac{d\zeta}{d\tau} = \eps(1-\xi^2)-\eta,
\end{eqnarray}
where $\eps = a^{-3/2}$. The system \eqref{eq:Michflowsca} has an equilibrium 
at $(-1,0,0)$ with eigenvalues $2\eps + \mathcal{O}(\eps^3)$ and $\pm i -\eps+\mathcal{O}(\eps^2)$, 
and an equilibrium at $(1,0,0)$ with eigenvalues $-2\eps + \mathcal{O}(\eps^3)$
and $\pm i +\eps+\mathcal{O}(\eps^2)$.

When $a$ grows, and therefore $\eps$ decreases, the measure of the set of
bounded orbits of \eqref{eq:Michflowsca} also grows. To study this limit,
introduce the variable $s=\xi+\zeta$ and cylindrical coordinates $(R,\theta)$
with $\eta = R\cos\theta$ and $\zeta = R\sin\theta$. Now when $\eps \ll 1$ and
$R$ is bounded from below, the dynamics is fast in $\theta$, namely
$\dot{\theta} = -1 +\mathcal{O}(\eps/R)$,  while it is slow in $s$. After
averaging over the fast angle, $R$ becomes also slow and the system reads 
\begin{eqnarray*}\label{eq:averagedMic}
\frac{ds}{d\tau} = \eps\left(1-s^2-\frac{R^2}{2}\right),\qquad%
\frac{dR}{d\tau} = \eps Rs.
\end{eqnarray*}
This system has the integral
\begin{eqnarray}\label{eq:MichInte}
	h=R^2\left(1-s^2-\frac{R^2}{4}\right) .
\end{eqnarray}
The domain of interest is $h\in[0,1]$. The level $h = 0$
contains the two saddle-foci of \eqref{eq:Michflowsca} at $(s,R)=(\pm 1,0)$. The
level $h = 1$ corresponds to an elliptic equilibrium $(s, R) =(0, \sqrt{2})$, which
approximates, as $\eps \to 0$, the intersection of an elliptic periodic orbit of
\eqref{eq:Michflowsca} with the Poincar\'e section $\{\zeta=0\}$. 
The level sets $h\in(0,1)$ are close to invariant circles on the Poincar\'e section 
of the flow of \eqref{eq:Michflowsca} \cite{DIKS13}. 
These correspond to two-dimensional invariant tori of \eqref{eq:Michflowsca}.  When $\eps \ll 1$,
the ratio of the two frequencies on the invariant tori is large.

\begin{figure}[ht]
\begin{center}
\begin{tabular}{ccc}
{\fns$a=10$}&{\fns$a=4.95$}&{\fns$a=3.7$}\\
\includegraphics[width = 0.3\textwidth]{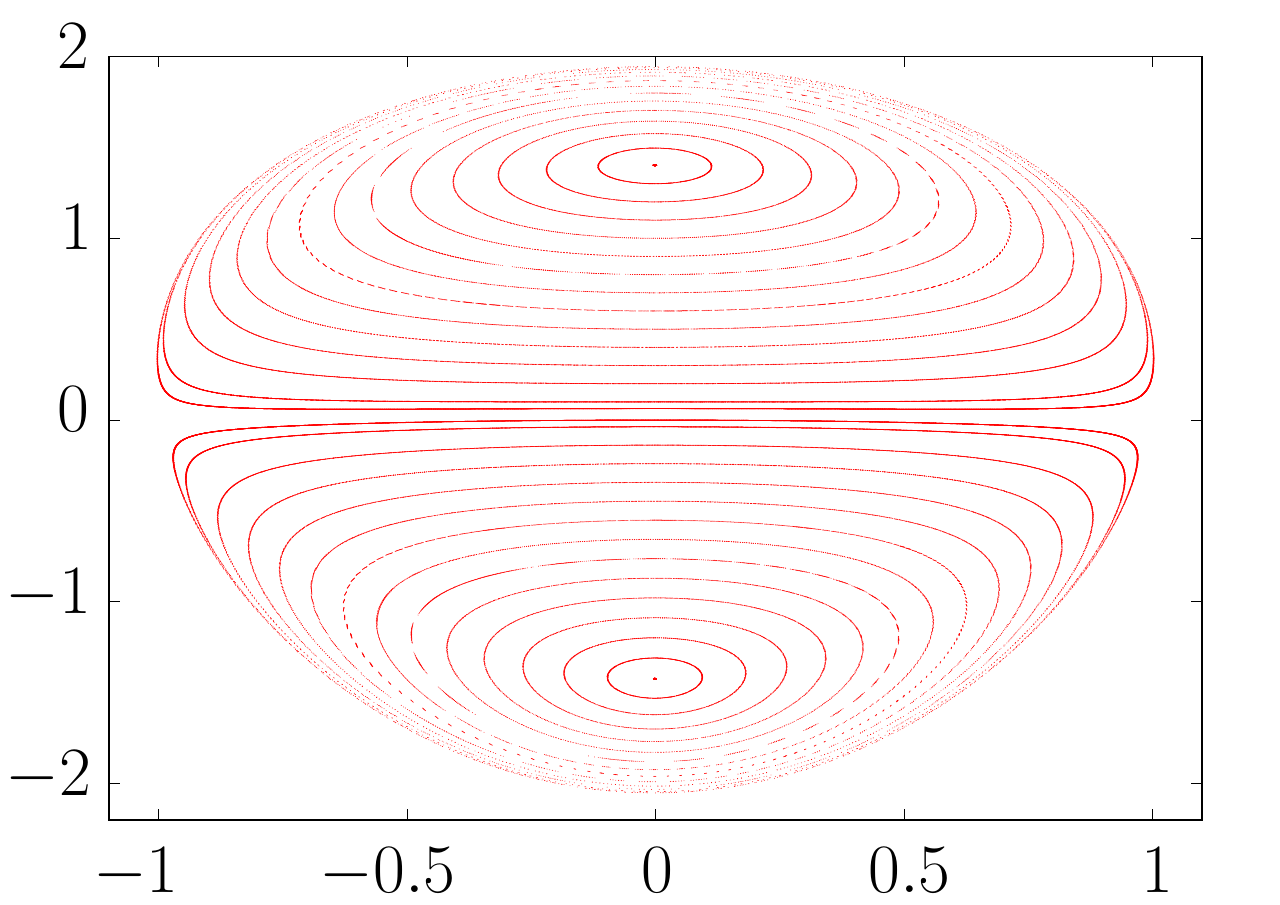}&
\includegraphics[width = 0.3\textwidth]{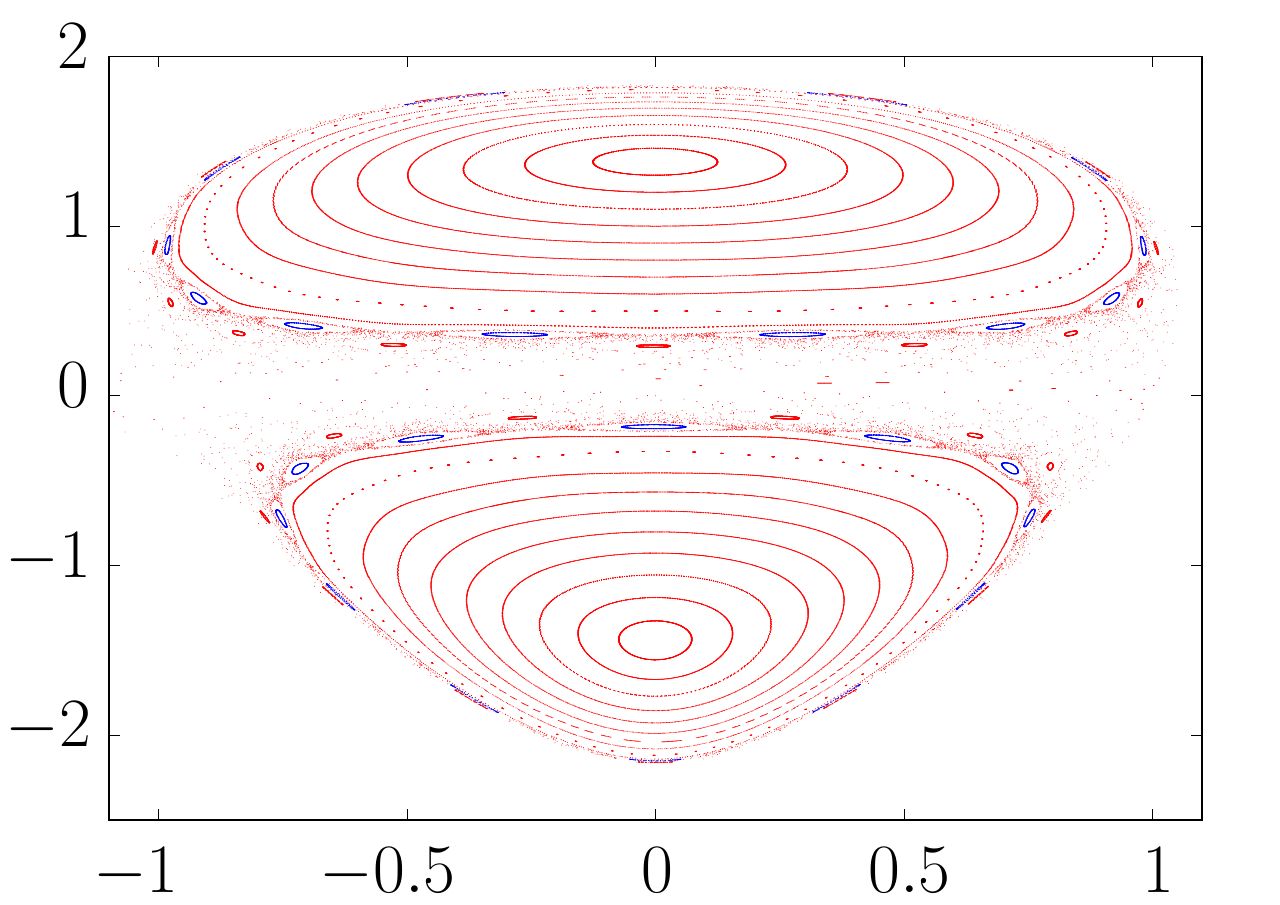}& 
\includegraphics[width = 0.3\textwidth]{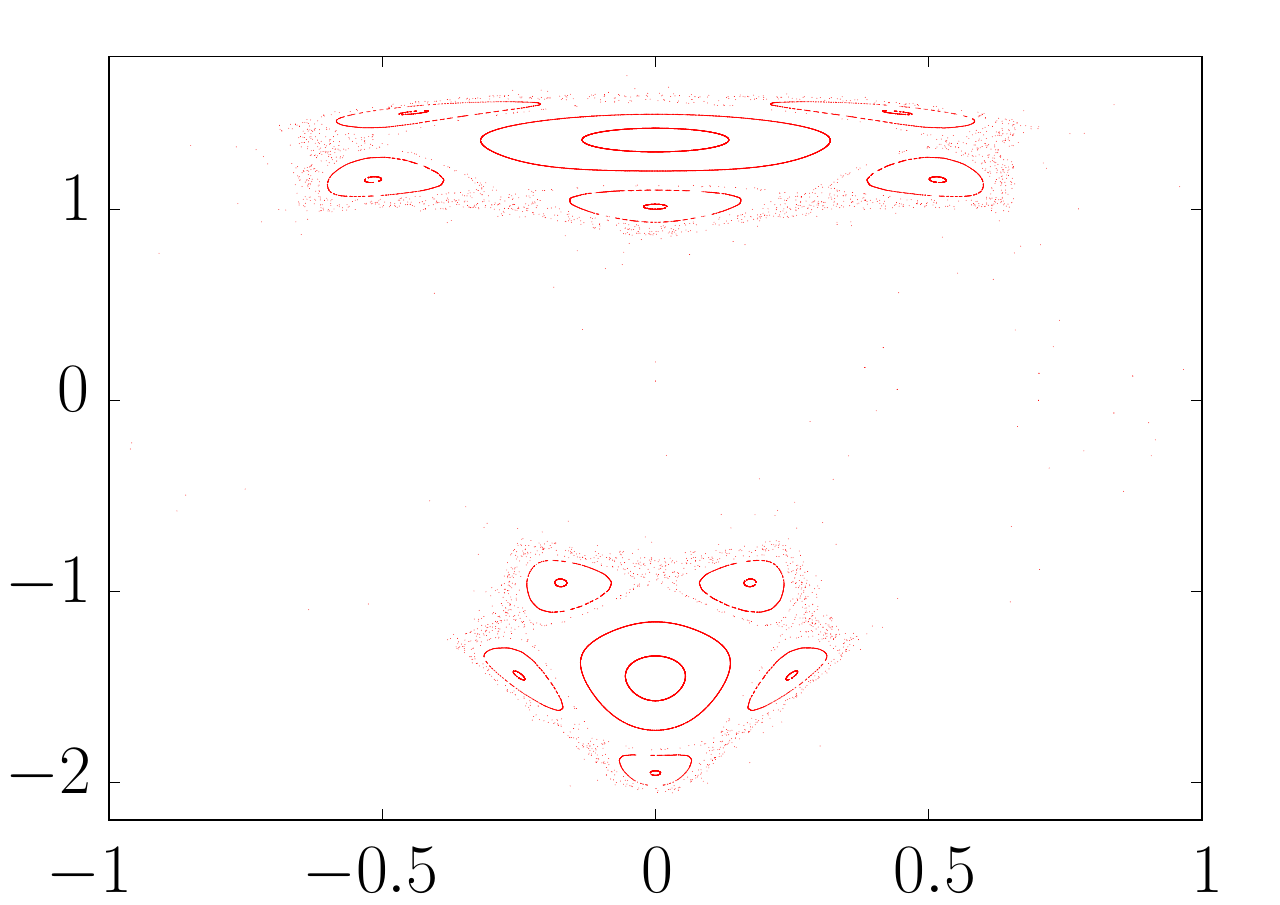}
\end{tabular}
\end{center} 
\caption{Slices $|\zeta| < \delta$ of trajectories of the map
\eqref{eq:Michelson_map}, in the $(\xi,\eta,\zeta)$ coordinates of
\eqref{eq:Michflowsca}, showing the rectangle $-1< \xi<1$,$-2<\eta<2$.  The
parameters are 
 $a = 10$,   $\vp = 0.1$,  $\delta = 0.001$  (left), 
 $a = 4.95$, $\vp = 0.01$, $\delta = 0.001$  (middle), and
 $a = 3.7$,  $\vp=0.001$,  $\delta = 0.0001$ (right).
}
\label{fig:Michmap-slice}
\end{figure}

More generally, the system \eqref{eq:Michflowsca} has two equilibria that are
saddle-foci: $Q^l=(-1,0,0)$ and $Q^r=(1,0,0)$ which have 1D invariant manifolds
$W^u(Q^l)$ and $W^s(Q^r)$ that nearly coincide as $\eps\to 0$. As $\eps$ tends to zero,
the 2D invariant manifolds $W^s(Q^l)$ and $W^u(Q^r)$ approach a spherical shell, that we refer
to as the \textit{bubble} \cite{BSV08a, BSV08b, BSV10a, BSV10b}. 
The bubble encloses a family of nested tori around a normally
elliptic invariant circle (see e.g., Fig.~\ref{fig:Michmap-slice} (a)) when
$\eps$ is small enough. If $\eps>0$ \eqref{eq:Michflowsca} is not integrable
and the 1D and 2D invariant manifolds no longer coincide \cite{BCS13, DIKS13,
BCS16a, BCS16b}. See \cite{DIKS13} for a detailed numerical study of the region
of bounded motion of \eqref{eq:Michelson_flow}.\footnote
{A movie of the evolution of
the flow with $a$ is at \url{http://www.maia.ub.es/dsg/moviehsn}.}

The quadratic map \eqref{eq:Michelson_map} is also not integrable. Fixing
$a>0$, the points $Q^l$ and $Q^r$ are saddle-foci when $\vp$ small enough
\cite{DM09}.  This occurs approximately when $a\vp^2 \in (0,4)$. More
precisely, if $\vp<\tfrac12$ it is sufficient to have $a\vp^2<3.87$ and if
$\vp<\tfrac14$ it is sufficient to have $a\vp^2<3.98$.  For these values of the
parameters some of the bubble structure of the flow is preserved. Namely, the
2D invariant manifolds of $Q^l$ and $Q^r$ (which do not coincide), 
bound a Cantor family of invariant tori that
enclose, for most values of the parameters $\vp$ and $a$, an elliptic invariant
circle \cite{DM09}.

When $\vp \ll 1$, the dynamics of \eqref{eq:Michelson_map} limits on that of
the ODEs \eqref{eq:Michelson_flow}. In Fig.~\ref{fig:Michmap-slice} we show,
in the $(\xi,\eta,\zeta)$ coordinates of \eqref{eq:Michflowsca}, the points
on some orbits which follow in the slice $|\zeta|<\delta \ll 1$. The orbits
shown in the three panels pass through the corresponding slice, moving
``up", $\zeta'>\zeta$, when $\eta\lesssim 0$ and ``down", $\zeta'<\zeta$,
if $\eta\gtrsim 0$. The boundary between the orbits going ``up" and ``down"
is $\eta=\eps(1-\xi^2)$.
The leftmost panel corresponds to $a=10$ for which the set of bounded orbits
resembles that of the integrable case discussed above: at $\xi\approx0$ and
$\eta \approx \pm \sqrt{2}$ we observe what seems two elliptic fixed points
that correspond to a transversally elliptic invariant circle.  The nested
invariant curves in the plot correspond to slices through invariant tori
surrounding this invariant circle. For the center panel, where $a=4.95$, the
outermost structure shows satellite islands and several unbounded orbits that
are temporarily trapped near the outermost torus.  A similar structure also
would appear in the left panel under a sufficiently high magnification. The
blue points correspond to the intersection of a satellite torus that performs
twelve complete turns around the $\xi$ axis before closing. Similar tori doing
9, 10, 11 and 13 turns before closing have also been easily detected. Finally
for the right panel, where $a=3.7$, the regular region has eroded, though there
are still some tori around the central invariant curve. Moreover, there appears
what seems to be a period-five elliptic invariant circle surrounded by tori
that are satellites of the central structure.

\section{A Volume-preserving map with accelerator modes}
\label{sect:model3D}

In this section we construct a 3D angle-action map with accelerator modes.  Our
goal is to study the stickiness of a bubble of regular orbits in an otherwise
seemingly fully chaotic phase space. Hence, we look for a family $f_\eps$ of
VPM of $\Tset^2\times\Rset$, that smoothly projects to a map $\tilde{f}_\eps$
on $\Tset^3$ under $\Pi$, recall \eqref{eq:lift}.

To construct our model, we choose $f_\eps$ so that it fulfills the
following three requirements (already sketched in \S\ref{sect:intro})
\begin{itemize}
	\item[{\bf R1}] The map has an integrable limit $\eps\to0$, where 
the phase space is foliated by horizontal rotational invariant tori (RIT)
$\{z={\rm const}\}$ and the restriction of the dynamics on each RIT is
topologically conjugate to a rigid rotation. Near this limit, some of these
tori should persist: a volume-preserving KAM theorem should apply \cite{CS89, Xia92}.
	\item[{\bf R2}] For $\eps=n\in\Nset_+$, the origin $P_+=(0,0,0)$ is a
fixed point of $\tilde{f}_n$, and for all $m\in\Zset$, $f_n^q(0,0,m) =
(0,0,m+nq)$. Hence, the origin is an FPAM, recall Def.~\ref{def:accelerator}.
	\item[{\bf R3}] Near the creation of the FPAM, the parameter
$\kappa_n=\eps-n$ unfolds a Hopf-one bifurcation. Hence, for $0 < \kappa_n \ll 1$, 
a small volume of regular orbits may exist near $P_+$. We
will define the family $f_\eps$ in such a way that its Taylor expansion around
$P_+$ for $\eps=n+\kappa_n,\,n\in\Nset_+$ is locally conjugate to a map in the
family $\Mic$ \eqref{eq:Michelson_map}, where the higher order terms (in $u$,
$v$, $w$) depend on $n$ in such a way that they tend to vanish as $n\to\infty$,
see Prop.~\ref{prop:dynloc} in \S\ref{sect:locdyn}.
\end{itemize} 

\subsection{Shearing functions}\label{sect:choiceshears} 

In this subsection we construct a concrete family of maps satisfying the
requirements {\bf R1}, {\bf R2} and {\bf R3} using the composition of three
shears \eqref{eq:tipus}.

The second and third requirements are achieved for the family
\begin{eqnarray}\label{eq:map3D}
	f_\eps:
		\left(\begin{array}{c}
			x\\y\\z
		\end{array}\right)
	\mapsto
		\left(\begin{array}{c}
			x'\\y'\\z'
		\end{array}\right)
	&=&
	\left(\begin{array}{l}
		x+\mu\sin(2\pi y)+\psi(z')\\
		y+\nu\sin(2\pi z')\\
		z+\eps\left(\cos(2\pi x)-\beta\sin(2\pi y)\right)
	\end{array}\right),
\end{eqnarray}
where $\mu, \nu, \beta$ are parameters. We assume that $\psi$ is a
degree-one circle map (i.e., $\psi(z+1) = z + \psi(z)$) that satisfies
\begin{equation}\label{eq:asspsi}
	\psi(0) = \psi'(0) = 0.
\end{equation}
To satisfy {\bf R1} the function $\psi(z)$ could simply be $z$ itself, and---as
we will show below---{\bf R2} is automatically fulfilled when $\psi(0)=0$. The
condition {\bf R3} requires, however, that the first derivative vanishes at the
location of the FPAM, see \S\ref{sect:locdyn}.

From the expression \eqref{eq:map3D} it is clear that 
$P_+=(0,0,0)$ and $P_-=(\tfrac12,0,0)$ are fixed points of the projection $\tilde{f}_n$. 
Under $f_n$, $P_+$ goes up by $n$ units and $P_-$ goes down by $n$ units
in $z$ upon each iterate (see also Rem.~\ref{rema:zeroflux}). After the Hopf-one bifurcation the point $P_+$
gives rise to a pair of FPAM, to be denoted by $P^{l,r}_+$ in
\S\ref{sect:locdyn}. It would be nice to have similar properties for $P_-$, i.e, for it to
give rise to a FPAM pair $P^{l,r}_-$ as well. A simple way to obtain this is by choosing
$\psi$ to be an odd function: $\psi(-z) = -\psi(z)$. This 
is not necessary to unfold the bifurcation, but it is simpler to have similar bubbles
created near $P_+$ and $P_-$, one going up and the other down.

To satisfy \eqref{eq:asspsi} and to have the odd character of $\psi$ we choose $\psi(z)-z$ to be an odd periodic
function given by the trigonometric polynomial
\begin{eqnarray}\label{eq:psi(z)}
	\psi(z)=z+\sum_{j=1}^7a_{j}\sin(2\pi j z).
\end{eqnarray}
The choice of the function $\psi$ above is justified in
App.~\ref{app:choicepsi}, where appropiate values for the Fourier
amplitudes, $a_j$, are also given by \eqref{eq:aValues}.

To ensure that \eqref{eq:map3D} fulfills {\bf R1} we can take
$$
	\mu=\eps\tilde{\mu}, \quad \tilde{\mu}=\Ocal(1).
$$
The point is that when $\eps = \mu = 0$ each horizontal two-torus $\{z={\rm
const}\}$ is invariant, and the dynamics on each torus is a rigid rotation with
rotation vector $\om=(\psi(z),\nu\sin(2\pi z))$. 

The first requirement is then satisfied if $f_\eps$ satisfies the hypotheses of
the volume-preserving KAM theorem \cite{CS89, Xia92}. This theorem is stated
for an analytic map of the form \eqref{eq:tipusmap}. Our model \eqref{eq:map3D}
can be written in this form upon taking
\begin{eqnarray*}\label{eq:triagih}
\Omega(z)         &=& (\psi(z), \nu\sin(2\pi z)),\\
\eps h_1(x,y,z) &=& \Omega_1(z')-\Omega_1(z)+\eps\tilde{\mu}\sin(2\pi y),\\
\eps h_2(x,y,z) &=& \Omega_2(z')-\Omega_2(z),\quad\mbox{and}\\ 
     h_3(x,y,z) &=&  \cos(2\pi x)-\beta\sin(2\pi y).
\end{eqnarray*}

In addition, we have to check if the following two necessary conditions hold
for $f_\eps$ \cite{Xia92}:
\begin{enumerate}
	\item {\it Intersection property}. The image under $f_\eps$ of any
homotopically non-trivial two-torus, sufficiently close to a horizontal torus
$\{z = \mbox{const}\}$, intersects itself. This is achieved because
$h_3(x,y,z)$ has zero average with respect to the angles $(x,y)$. 
	\item {\it Nondegeneracy condition}. There exists a $k\in\Nset$, such
that the frequency map satisfies a twist-like, nondegeneracy condition:
\begin{eqnarray}\label{eq:nondeg}
{\rm rank}
\left(\begin{array}{cc}
\Omega_1'(z)      & \Omega_2'(z)\\
\vdots       & \vdots\\
\Omega_1^{(k)}(z) & \Omega_2^{(k)}(z)
\end{array}\right)=2.
\end{eqnarray}
\end{enumerate}
If $|\eps| \ll 1$, $\mu=\Ocal(\eps)$, and $\psi(z)$ is chosen to
satisfy \eqref{eq:nondeg}, KAM theory implies that $f_\eps$ will have
a Cantor set of RIT that are deformations of the horizontal tori that exist for
$\eps=0$. 

\begin{rema} By contrast with the case of symplectic maps, since the number of
actions is less than the number of angles ($l < d$), the frequency map $\Omega:
\Rset^l \to \Tset^d$ cannot be surjective. Hence one cannot assure the
persistence of a RIT with prescribed frequencies.  Thus KAM theory does not
guarantee the persistence of a torus with a given rotation vector, only that
there are many tori when $\eps \ll 1$. 
\end{rema}

\begin{rema} \label{rema:zeroflux} For the map \eqref{eq:tipusmap}, the condition that $h_3$ has zero
average means that there is zero net volume flux through any RIT. This
condition implies the intersection property.  For the map \eqref{eq:map3D},
this condition also implies that for each FPAM with positive acceleration, e.g.
$P_+$, there is another with negative acceleration. In our case, the
corresponding downwards FPAM is located at $P_-=(\tfrac12,0,0)$.
\end{rema}

The nondegeneracy condition \eqref{eq:nondeg} may have a different minimal
value of $k$ in different ranges of $z$. For example, for $f_\eps$,
\eqref{eq:nondeg} does not apply for $k=2$ at $z=0$ since $\psi'(0) = \psi''(0)
= 0$.  However, it will hold for $k=3$ so long as $\psi^{(3)}(0) \neq 0$.  This
may happen for other values of $z$ depending on the choice of $\psi(z)$.
Indeed, since $\psi(z)$ is odd, \eqref{eq:nondeg} for $k=2$ is also violated at
$z=\tfrac12$.  Consequently, we expect that there will be more prominent chaotic
zones near $\{z=0\}$ and $\{z=\tfrac12\}$ for small values of $\eps>0$. For the
choice \eqref{eq:psi(z)} with the coefficients
\eqref{eq:aValues}, the condition \eqref{eq:nondeg} is violated at ten
additional values of $z\in(0,1)$ for $k=2$, but one can check that it does hold
for $k=3$ at all of these points. 

To verify that \eqref{eq:map3D} satisfies {\bf R2}, we can compute its fixed
points and accelerator modes. For any values of the parameters, there are four
fixed points located at $(\tfrac14,0,0)$, $(\tfrac14,\tfrac12,0)$,
$(\tfrac34,0,0)$, and $(\tfrac34,\tfrac12,0)$. Since the map preserves volume,
all of these are generically unstable: the product of the three multipliers of
$Df_\eps$ is one, $\lambda_1 \lambda_2 \lambda_3 =1$. So, unless all three have
modulus one, there will be at least one unstable multiplier.  Additional fixed
points correspond to accelerator modes. The following Lemma is proved in
App.~\ref{app:FPAM}.

\begin{lema}\label{lema:FPAM} Suppose
that 
\begin{equation}\label{eq:munuRange}
	0<|\mu|<\tfrac12, \qquad 0<|\nu|<\tfrac12.
\end{equation}
Then for each $\eps=n\in\Nset_+$ $f_\eps$ has a Hopf-one bifurcation that
creates four FPAM. Two of these, $P_+=(0,0,0)$ and $Q_+=(0,\tfrac12,0)$,
accelerate upwards, and two, $P_-=(\tfrac12,0,0)$ and
$Q_-=(\tfrac12,\tfrac12,0)$, accelerate downwards.
\end{lema}

Finally, we note that the map $f_\eps$ commutes with the involution $R$: 
$f_\eps \circ R = R \circ f_\eps$, where $R$ is given by 
\begin{equation}\label{eq:reversor}
	R(x,y,z) = (\tfrac12 -x, -y, -z).
\end{equation}
Indeed, this follows for any map of the form \eqref{eq:tipus} when the shears
are odd about the point $(\frac14, 0,0)$, which is a fixed point of $R$.
In particular $R(P^l_+)=P^r_-$ and $R(P^r_+)=P^l_-$. Also the manifolds
associated to the $P^{r,l}_-$ are obtained under the symmetry $R$ from the
manifolds of $P^{l,r}_+$. See \S\ref{subsect:geom} for details.

For the remainder of the paper, we will not vary $\mu$ with $\eps$, but will
return to the form \eqref{eq:map3D} for a fixed small value of $\mu$. We think
of $\eps$ as the primary parameter, and take $(\mu, \nu, \beta)$ as ``fixed". 

\subsection{Local dynamics near an accelerator mode}\label{sect:locdyn}

In this section we study the local dynamics around the FPAM of $f_\eps$ \eqref{eq:map3D} when
$\eps$ is near $n\in\Nset_+$. This is done by expanding about the FPAM
to quadratic order and rescaling the variables.

To motivate the scaling, consider for example, the dynamics around
$P_+=(0,0,0)$. Let $\eps=n+\kappa_n$, where $\kappa_n>0$ is small. Then $P_+$
bifurcates into a pair of new FPAM located at $P_+^{l,r} = (x^{l,r},0,0)$ where
\eqref{eq:map3D} implies that $x^{l,r}$ must satisfy $(n+\kappa_n)\cos(2\pi x^{l,r}) = n$.
When $\kappa_n$ is small, this implies
\begin{equation}\label{eq:xlr}
x^{l,r}  = 
\mp\frac{1}{\pi}
\sqrt{\frac{\kappa_n}{2n}} + \mathcal{O}(\kappa_n).
\end{equation} 
This scaling motivates the introduction of a new parameter $\delta=n\kappa_n$
and of the scaled phase variables $n(x,y,z)$, so that the
distance between the new FPAM becomes $\Ocal(\sqrt{\delta})$.

\begin{prop}\label{prop:dynloc} 
Given $\mu, \nu, \beta$, let $\eps=n+\delta/n$ and $P$ be any of the accelerator modes of Lemma~\ref{lema:FPAM}.
Thus $\delta/n$ measures the distance from the birth of $P$. Define new phase variables
$(\xi,\eta,\zeta) = n\left((x,y,z)-P\right)$, and let
$f^*_\delta(\xi,\eta,\zeta)$ be the projected map $\tilde{f}_{n+\delta/n}$ in the new variables.
Then the following holds.
\begin{enumerate}
	\item The Taylor expansion of $f_{\delta}^*$ around the origin  can be written
as $f^*_\delta = L + \Ocal(n^{-1})$, where $L$ is a quadratic volume-preserving
map.
	\item An additional normalization $(u,v,w) = (\alpha_\xi
\xi,\alpha_\eta \eta,\alpha_\zeta \zeta)$ conjugates $L$ to the Michelson map
\eqref{eq:Michelson_map} for suitable parameters $\vp$ and $a$. 
\end{enumerate} 
\end{prop}

\noindent{\it Proof.} For the moment, let us restrict ourselves to the dynamics
around $P_+$. In the variables $(\xi,\eta,\zeta) = n(x,y,z)$, map $f_\delta^*$
becomes
\begin{eqnarray}\label{eq:fepsescalada}
\left(\begin{array}{c} \xi'\\\eta'\\\zeta' \end{array}\right)
=
\left(\begin{array}{c}
\xi   + n\mu\sin\left(2\pi\frac{\eta}{n}\right)+n\psi\left(\frac{\zeta'}{n}\right)\\
\eta  + n\nu\sin\left(2\pi\frac{\zeta'}{n}\right)\\
\zeta + n\left(\left(n+\frac{\delta}{n}\right)\left(\cos\left(2\pi\frac{\xi}{n}\right)-\beta\sin\left(2\pi\frac{\eta}{n}\right)\right)-n\right)
\end{array}\right),
\end{eqnarray}
where $-n$ in the third component is due to the projection to the torus.
Expanding around $(0,0,0)$ gives
$$
f^*_\delta =L_{\delta, \beta}+\Ocal(n^{-1}),
$$ 
where
\begin{eqnarray}\label{eq:fepsescaladalim}
L_{\delta, \beta}:
\left(\begin{array}{c}
\xi'\\\eta'\\\zeta'
\end{array}\right)
=
\left(\begin{array}{ccl}
\xi   &+& 2\pi\mu\eta \\ 
\eta  &+& 2\pi\nu\zeta' \\
\zeta &+& \delta - 2\pi^2 \xi^2 - 2\pi \beta n \eta
\end{array}\right).
\end{eqnarray}
Note that $n$ has disappeared, except for the last term, proportional to $\beta n$.

The same procedure can be applied to the remaining three FPAM, but one has to
take into account some changes of sign due to expanding the trigonometric
functions around $\pi$ instead of $0$, and the fact that $P_-$ and $Q_-$ jump
downwards. Table~\ref{tab:scalings} summarizes the scalings and gives the form
of $L$ one obtains after this procedure. Note that the only difference in the 
final form is that $\beta \to -\beta$ for the $Q_\pm$ FPAM.

\begin{table}[ht]
\begin{center}
\begin{tabular}{l|l|l}
FPAM & $(\xi,\eta,\zeta)$ & Map\\
\hline
$P_+ = (0,0,0)$     &$n(x,y,z)$ & $L_{\delta, \beta}$\\
$P_- = (\tfrac12,0,0)$   &$n(\tfrac12-x,-y,-z)$  & $L_{\delta, \beta}$\\
$Q_+ = (0,\tfrac12,0)$   &$n(-x,y-\tfrac12,z)$   & $L_{\delta,-\beta}$\\
$Q_- = (\tfrac12,\tfrac12,0)$ &$n(x-\tfrac12,\tfrac12-y,-z)$ & $L_{\delta,-\beta}$
\end{tabular}
\end{center}
\caption{Scalings to obtain the quadratic map $L$ near an FPAM.}
\label{tab:scalings}
\end{table}
Applying the additional normalization $(u,v,w) = (\alpha_\xi \xi, \alpha_\eta \eta,
\alpha_\zeta \zeta)$ to \eqref{eq:fepsescaladalim} shows that $L_{\delta, \beta} \simeq
\Mic$, the Michelson map \eqref{eq:Michelson_map}, if we choose

\begin{eqnarray}\label{eq:escalats_3D}
&
\begin{array}{ccccccccccc}
\alpha_\xi   &=& \displaystyle\pi \left(\frac{2}{\delta}\right)^{\frac12},          &\,&
\alpha_\eta  &=& \displaystyle\pi \left(\frac{4\mu^2}{\delta^2\nu}\right)^{\frac13},&\,&
\alpha_\zeta &=& \displaystyle\pi \left(\frac{32\mu^2\nu^2}{\delta^5}\right)^{\frac16},
\end{array}
&\\
\label{eq:parametres_3D}
&
\begin{array}{ccccccc}
\vp   &=&\pi\left(32\mu^2\nu^2\delta\right)^{\frac16},&\,&a &=&\displaystyle
\beta n\left(\frac{2\nu}{\delta\mu^2}\right)^{\frac13}.
\end{array}
&
\end{eqnarray}
These expressions are the same for the other fixed points except that for
$Q_\pm$, $a$ changes sign since, by Tbl.~\ref{tab:scalings}, $\beta \to -\beta$.
\fin

\begin{rema}\label{rema:rema1-3D}
There are some important aspects of the local form that are worth noting:
\begin{itemize}
	\item The fixed points of \eqref{eq:fepsescaladalim}, at $(\pm
\sqrt{\tfrac{\delta}{2\pi^2}},0,0)$, collide as $\delta\to0$. 
	\item For fixed $\beta$, the parameter $a$ as given in
\eqref{eq:parametres_3D} grows linearly with $n$. Recall, from
\S\ref{sect:hopfone}, that a bubble of stability for $\Mic$ appears when
$a\vp^2 = 4\pi^2 \beta n\nu \in(0,4)$.  Hence, for the one-parameter family
$f_\eps$, we can only expect to detect a finite number of such stability
regions, those born at $\eps = n<(\beta\pi^2\nu)^{-1}$.  
	\item A bubble of stability occurs near  $P_\pm$ when $\beta\nu >0$,
but since the sign of $a$ in \eqref{eq:parametres_3D} changes for $Q_\pm$, the
bubble will occur near $Q_\pm$ when $\beta\nu <0$. Hence the requirement {\bf
R3} is satisfied. 
\end{itemize}
\end{rema}

Proposition~\ref{prop:dynloc} implies that $\Mic$ encodes the local dynamics
near an FPAM under the proper scaling. To do this, we think of $f^*_\eps$ as a
two-parameter family $f^*_{\eps,\beta}$.  A final scaling of the parameter
$\beta$ implies the following.

\begin{corol} For given $\mu, \nu$  let $\beta_n=\beta/n$ and
$\eps_n=n+\delta/n$ for fixed $\beta$ and $\delta$.  Then there is a ball
around $P_\pm$ ($Q_\pm$) inside of which the Taylor expansion of
$f^*_{\eps_n,\beta_n}$ converges, as $n\to\infty$, to a map that is conjugate
to $\Mic$ ($M_{\vp,-a}$), where
$\vp=\pi(32\mu^2\nu^2\delta)^{1/6}$ and
$a=\beta\left({2\nu}/{\delta\mu^2}\right)^{1/3}$.
\end{corol}

\section{Diffusion in the presence of a bubble: a case study}
\label{sect:diffusion3D}

In this section we study the diffusive properties of chaotic orbits of
\eqref{eq:map3D} when there is a bubble of stable orbits near some of the FPAM,
see App.~\ref{app:FPAM}. To this end, we perform numerical simulations for $0< \eps-1 \ll 1$. 

\subsection{Choosing parameters}\label{sect:choiceparam}

We use the function $\psi(z)$ introduced in App.~\ref{app:choicepsi} and choose
values of the parameters $\mu, \nu$ and $\beta$ of $f_\eps$ in \eqref{eq:map3D}
so that
\begin{enumerate}
	\item For $\eps-1>0$ and small the local map $\Mic$
\eqref{eq:Michelson_map} around $P_\pm$, satisfies $a\vp^2=4\beta\pi^2\nu\in(0,4)$.
	\item The critical parameter value, $\eps_{\rm crit}$, at which the
last RIT of $f_\eps$ is destroyed is as large as possible.
\end{enumerate}
The first requirement is a necessary condition to ensure that there is a region
of regular motion near the FPAM $P_\pm$. The second requirement, ensures that
the map is not too chaotic. Note that the value $\eps_{\rm crit}$ is analogous
to Greene's critical value 
for Chirikov's standard map \eqref{eq:stdmap} \cite{Gre79}. Such values have been
found for VPM in \cite{Mei12, FM13}. 

After an exploration of the dynamics for various parameters, we choose
\begin{eqnarray}\label{eq:parameters}
\mu = 0.01,\qquad \nu = 0.24, \quad \mbox{and}\quad \beta = 0.12.
\end{eqnarray}
For this choice, $a \vp^2 \approx 1.137 n$, so we only expect
to detect a region of regular motion around $P_\pm$ for $n<4$, recall
Rem.~\ref{rema:rema1-3D}. 

For the parameters \eqref{eq:parameters}, we conjecture that\footnote{
This critical value is not too far from Greene's critical value $k_{\rm crit} \approx 0.971635/(2\pi) \approx 0.154641$.
}
$\eps_{\rm crit}\in(0.093,0.094)$. To determine this, we iterated a set of initial
conditions in $\Tset^2\times[0,1]$ for $T = 2\cdot10^{7}$.  Each initial
condition was classified first as either escaping or non-escaping from
$z\in[0,1]$.  Those that did not escape were classified as either chaotic or
regular using an approximation of the Lyapunov exponent. If this approximation
was small, so that the orbit could be considered to be regular, we checked
whether it could be on a RIT by looking to see if its $(x,y)$ projection
completely filled all the pixels on a $400\times400$ grid.  

\subsection{Regular region around the accelerator mode}

\begin{figure}[ht]
\begin{center}
\begin{tabular}{cc}
\includegraphics[width = 0.45\textwidth]{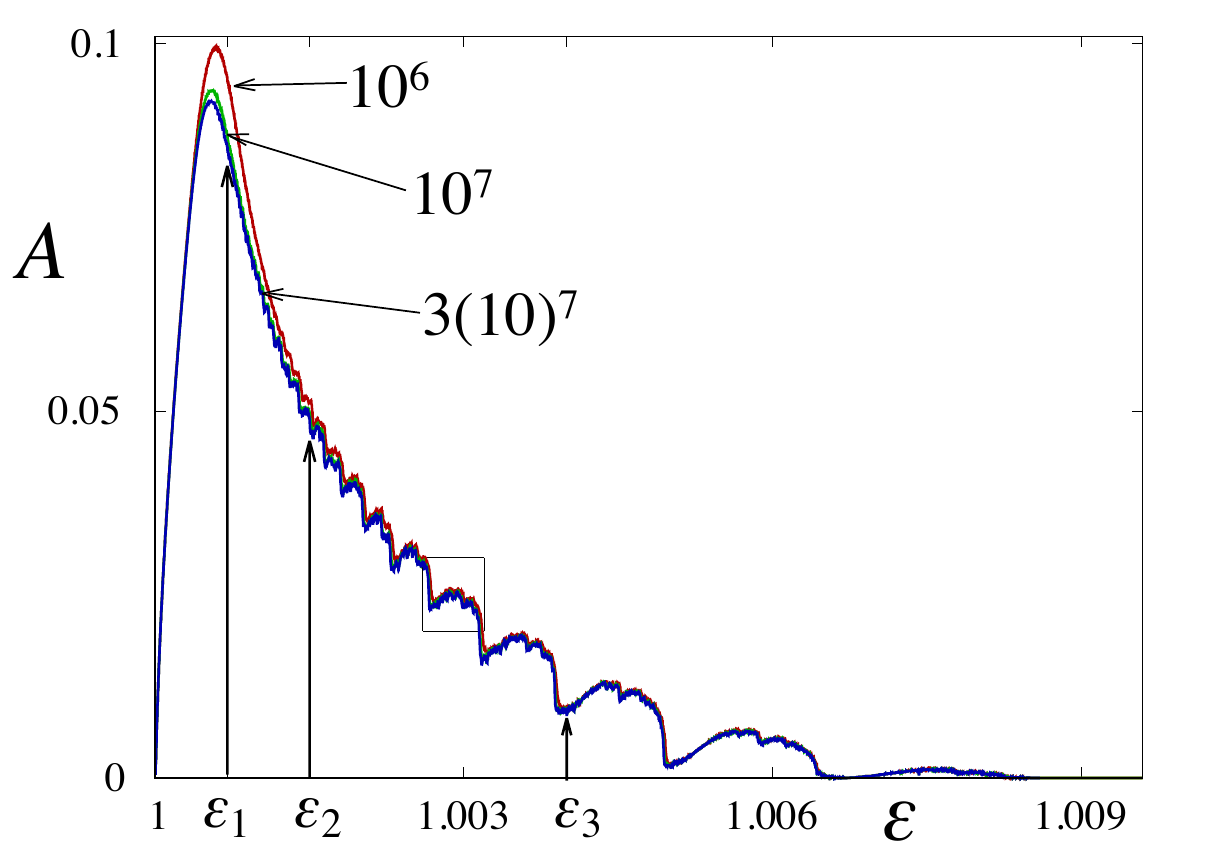}&
\includegraphics[width = 0.45\textwidth]{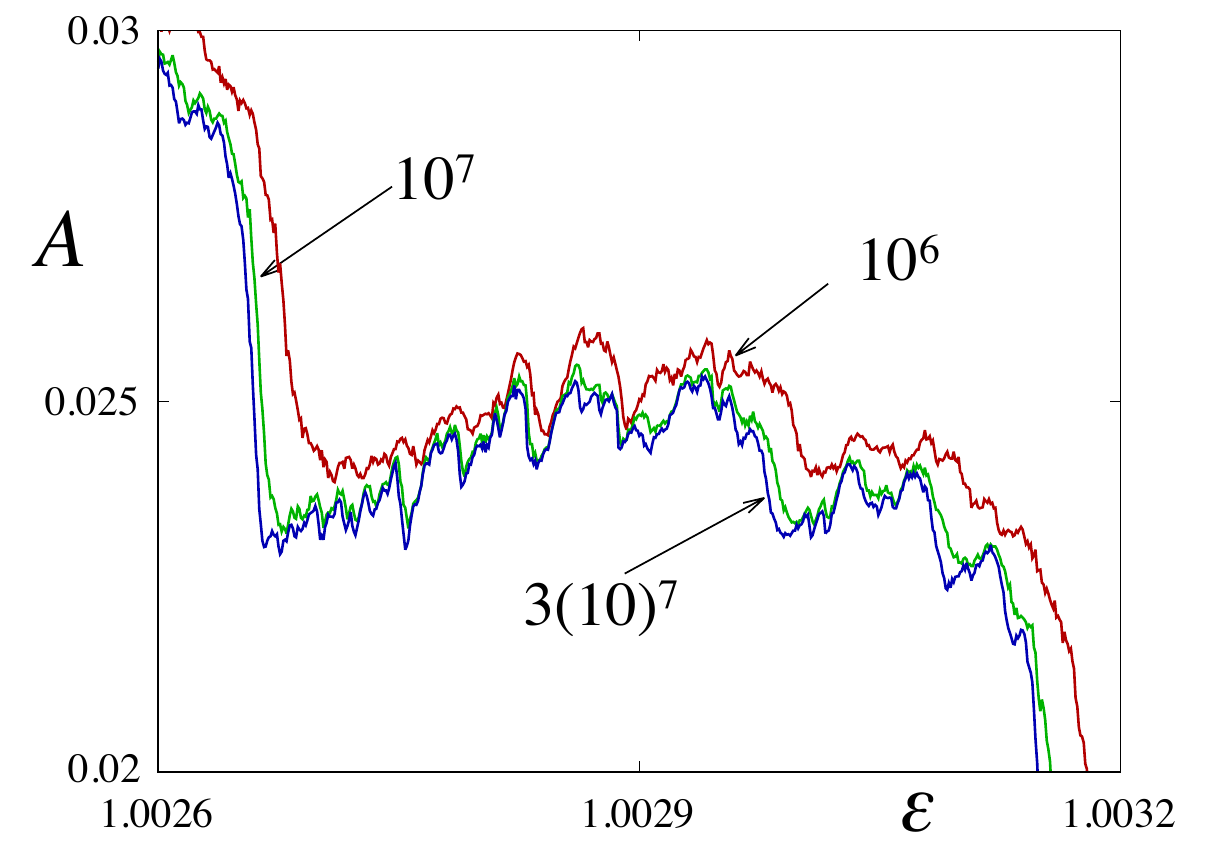}
\end{tabular}
\end{center}
\caption{The fraction of bounded orbits around the accelerator mode $P_+$ of
$f_\eps$  as a function of $\eps$ for parameters \eqref{eq:parameters}.
Initial conditions are chosen with $z=0$ in the box $(x,y)
\in[-0.024,0.024]\times[-0.12,0]$.  The three curves correspond to different
maximal number of iterates $T_{\rm max}$,  as labelled.  Left: the bounded
fraction for  $\eps=1+\kappa_1$, for $\kappa_1 \in [10^{-6},0.0096]$ in steps
of $10^{-6}$.  The labelled values $\eps_1 = 1.0007$, $\eps_2=1.0015$ and
$\eps_3=1.004$ are studied in \S\ref{sect:experiments}.  Right: magnification
near $\eps = 1.003$ of the box in the left figure.}
\label{fig:bounded_actualAM}
\end{figure}

We focus on the effect of the FPAM that appear for $\eps=1$, since they are
expected to have the largest bubble. Figure ~\ref{fig:bounded_actualAM} shows
the relative measure of bounded orbits near $P_+$ that start in the half-plane
$z=0,\,y\leq0$.  We considered a $400\times360$ grid in $(x,y)
\in[-0.024,0.024] \times[-0.12,0]$.  This range is chosen accordingly to the
position of the fixed points of $\tilde{f}_\eps$ that bifurcate from the origin
at $\eps=1$.  We iterate the centers of the grid cells up to a time $T_{\rm
max}$, and declare that the orbit escapes from the bubble if at any time
$\max(|x|,|y|,|z|) > 0.25$. The left panel of the plot shows values of $\eps$
over the full range where a stable accelerator mode with $n = 1$ is detected.
The fraction of bounded orbits exhibits a number of sudden decreases, and an
enlargement of some of these are shown in the right panel of the figure.  These
drops in bounded area correspond to the breakdown of an outermost invariant
two-torus that allows previously confined motion to escape from the bubble.

\subsection{Diffusion in the chaotic zone: expectations}\label{sect:expectations}

After the breakdown of the last RIT near $\eps_{\rm crit}\approx0.094$, the
phase space seems to become much more chaotic. In particular, for $0.2<\eps<1$
we have numerically checked that any regular component in the phase space is
below pixel size ($1/400^2$ squared units in $\Tset^2$).  For $\eps$ in a
subinterval of $[1,1.009]$ one detects the presence of a bubble of stability
around $P_{\pm}$, recall Fig.~\ref{fig:bounded_actualAM}. In this section we
investigate the diffusion in the $z$ variable for $\eps\in[0.2,1.8]$.

Outside the range in $\eps$ where the accelerator-mode bubble appears, we
expect an exponential decay of correlations giving ``normal" diffusion in the
action variable $z$, namely, that the standard deviation after $T$ iterates
\begin{equation}\label{eq:sigmaT}
	\sigma_T = \left(\left<(z_T - z_0)^2\right> - \left< z_T-z_0\right>^2 \right)^{\frac12} \sim T^\chi,
\end{equation}
where $\chi = \tfrac12$, so that the limit
\begin{eqnarray}\label{eq:difcoef}
D =\lim_{T\to\infty}\frac{\sigma_T^2}{2T},
\end{eqnarray}
exists. Here $\left<\cdot\right>$ stands for the  average over an ensemble of
initial conditions $(x_0,y_0,z_0)$, which we usually take to be uniform on some
domain of $\Tset^3$ outside bubbles of stability, and $(x_T,y_T,z_T) =
f_\eps^T(x_0,y_0,z_0)$. The one-step coefficient, known as the {\sl
quasilinear approximation}, can be easily evaluated as
\begin{eqnarray}\label{eq:quasilinear3D}
  D_{ql}  =\frac{1}{2}\left<(z' - z)^2\right>  = \int_{\Tset^2}(z'-z)^2  =
\frac{\eps^2}{4}(1 + \beta^2),
\end{eqnarray}
using \eqref{eq:map3D}.

The behavior of the action diffusion when there is a bubble, e.g., for
$\eps\in[1,1.009]$, can be expected to be very different. Indeed as was
discussed in \S\ref{sect:apm}, the hierarchical island-around-island structure
of the 2D case gives rise to a power-law behavior of the trapping time
distribution \cite{MO86}, which, in turn, gives rise to anomalous diffusion
\cite{AK08}. However for the 3D case, the way that tori in a bubble are
organized by their rotation vectors is not known, so we do not have the ability
to create a model similar to the 2D one.

\subsection{Numerical experiments}\label{sect:experiments}

In this section we describe the results of the numerical experiments for
diffusion and trapping statistics.  In \S\ref{sect:overallsigma} we will show
that the presence of accelerator-mode orbits gives rise to anomalous diffusion
of the action. In \S\ref{sect:algdecaytrap} we show that the trapping
statistics appears to have power-law decay $\Pcal_\eps(t)\sim t^{-b}$, $b>2$.
Both of these results are consistent with the 2D case \cite{Mei92,Mei97}.

In order to avoid choosing initial conditions inside a bubble, we consider them on a
fundamental domain of the right branch of the 1D unstable manifold of the fixed
point $(\tfrac34,0,0)$. When $\eps = \Ocal(1)$ and $\mu, \nu$ and $\beta$ as
given in \eqref{eq:parameters}, this point is a saddle with a 1D unstable
manifold and a 2D stable manifold. We choose $N=10^6$ to $10^7$ initial
conditions on $W^u(\tfrac34,0,0)$, logarithmically equispaced over a distance
interval $[10^{-9},10^{-8}]$ from the fixed point. 

Each initial condition was iterated between $10^8$ and $10^{11}$ times,
depending on the observed behavior, and we compute the following two
observables:
\begin{enumerate}
	\item {\sl The standard deviation}.  Anomalous diffusion of the action
is detected by examining the growth rate of $\sigma_T$, \eqref{eq:sigmaT}. In a
phase space that is seemingly fully chaotic and has no accelerator modes, one
expects the limit \eqref{eq:difcoef} to exist and that $D$ should be near the
quasilinear value \eqref{eq:quasilinear3D}.  When there are accelerator modes
one expects a faster growth so that the limit \eqref{eq:difcoef} does not
exist.  
	\item {\sl The trapping statistics}. We kept track of the number of
consecutive iterates that an orbit remains close to a bubble, i.e.,  in the
union $\Wcal = \Wcal_+\cup\Wcal_-$ of neighborhoods of $P_+$ and $P_-$.  For
most of cases, the neighborhoods
\begin{equation}\label{eq:stickyregion}
\begin{split}
	\Wcal_+ &=\{(x,y,z)\;:\;|x|\leq0.024,|y|\leq0.12,|z|\leq0.08\},\\
	\Wcal_- &=\{(x,y,z)\;:\;|x-\tfrac12|\leq0.024,|y|\leq0.12,|z|\leq0.08\},
\end{split}
\end{equation}
appear to completely contain the bubbles; however, we modify these regions
slightly in \S\ref{subsect:stat}.  Note that the set
$\Wcal_+\cap\{z=0,\,y\leq0\}$ was used in Fig.~\ref{fig:bounded_actualAM}.  The
probability of having a stay of exactly length $t$ near the bubbles is
\begin{equation}\label{eq:trappingprob}
\Pcal_\eps(t) = {\rm Prob}\left(
	 		(x_j,y_j,z_j\mbox{ mod }1) \left\{
				\begin{array}{ll}
					\in\Wcal,     & j\in[i,\ldots,i+t],\\
	 				\notin\Wcal,  & j \in\{i-1,i+t+1\}
			\end{array}\right\} : i \in [1,T-t] \right).
\end{equation}
This is the analogue of the trapping statistic \eqref{eq:trappingstat}
used in the area-preserving case.
\end{enumerate}

We computed $\Pcal_\eps(t)$ for an orbit of length $T = 2^{26.6} \approx 10^8$
by partitioning this interval into subintervals that are logarithmically
equispaced, i.e., $I_i=[2^{0.1 i},2^{0.1(i+1)})$ for $i$ up to $265$.  We declare an orbit
to be ``trapped" around a bubble if it remains in $\Wcal$ for at
least $t_0 = 128$ consecutive iterates, so we start with $i=70$, corresponding
to this shortest trapping segment.

A histogram is constructed for the number of trapped orbit segments in $\Wcal$
of length $t \in I_i$. Normalizing this gives the probability, $\Pcal_\eps$ for
$t = 2^{0.1(i+1/2)}$, in the logarithmic middle of $I_i$.
 
\subsubsection{Normal and anomalous diffusion}\label{sect:overallsigma} 

\begin{figure}[ht]
\begin{center}
\begin{tabular}{ccc}
\hspace{-0.25cm}%
\includegraphics[width = 0.35\textwidth]{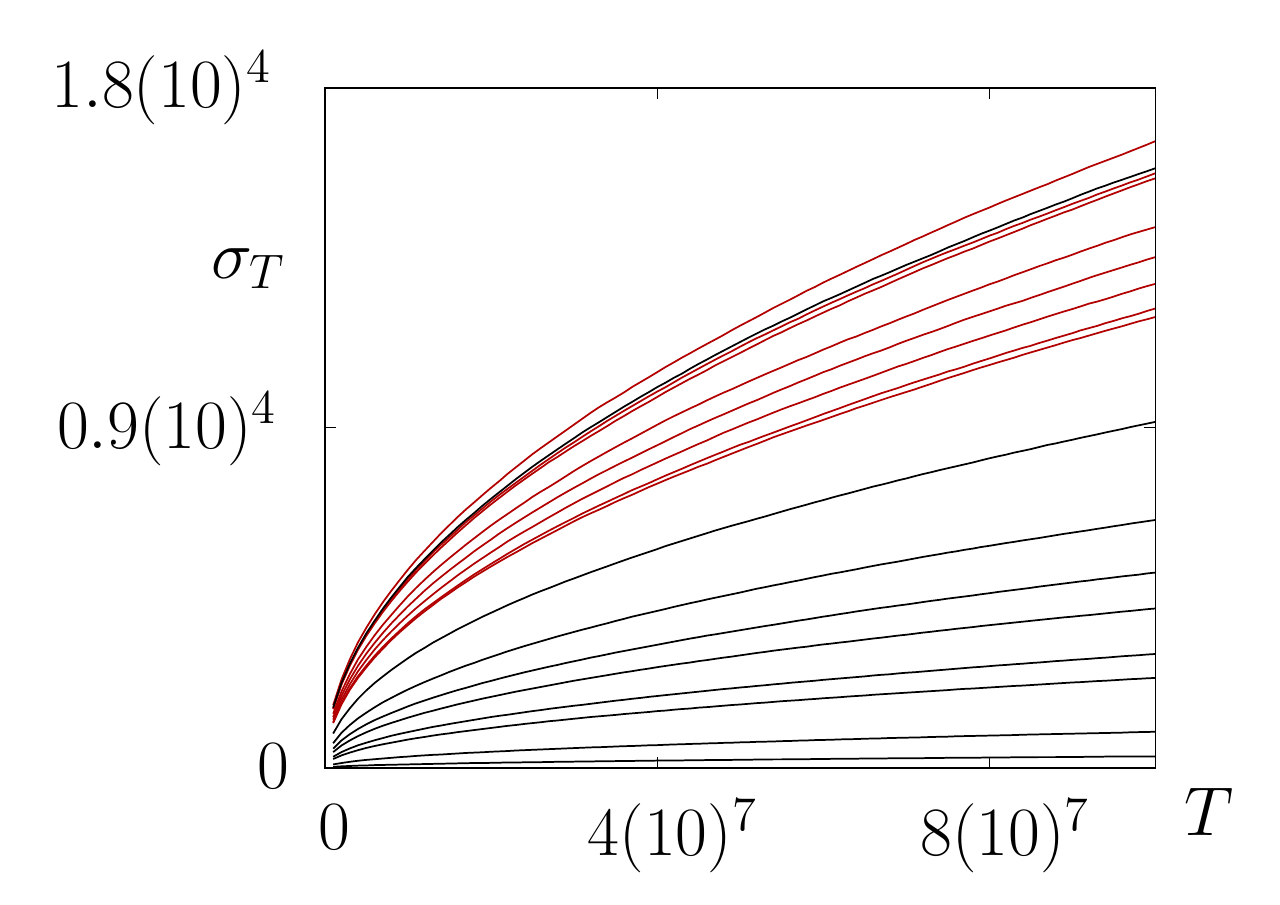}&
\hspace{-0.75cm}%
\includegraphics[width = 0.35\textwidth]{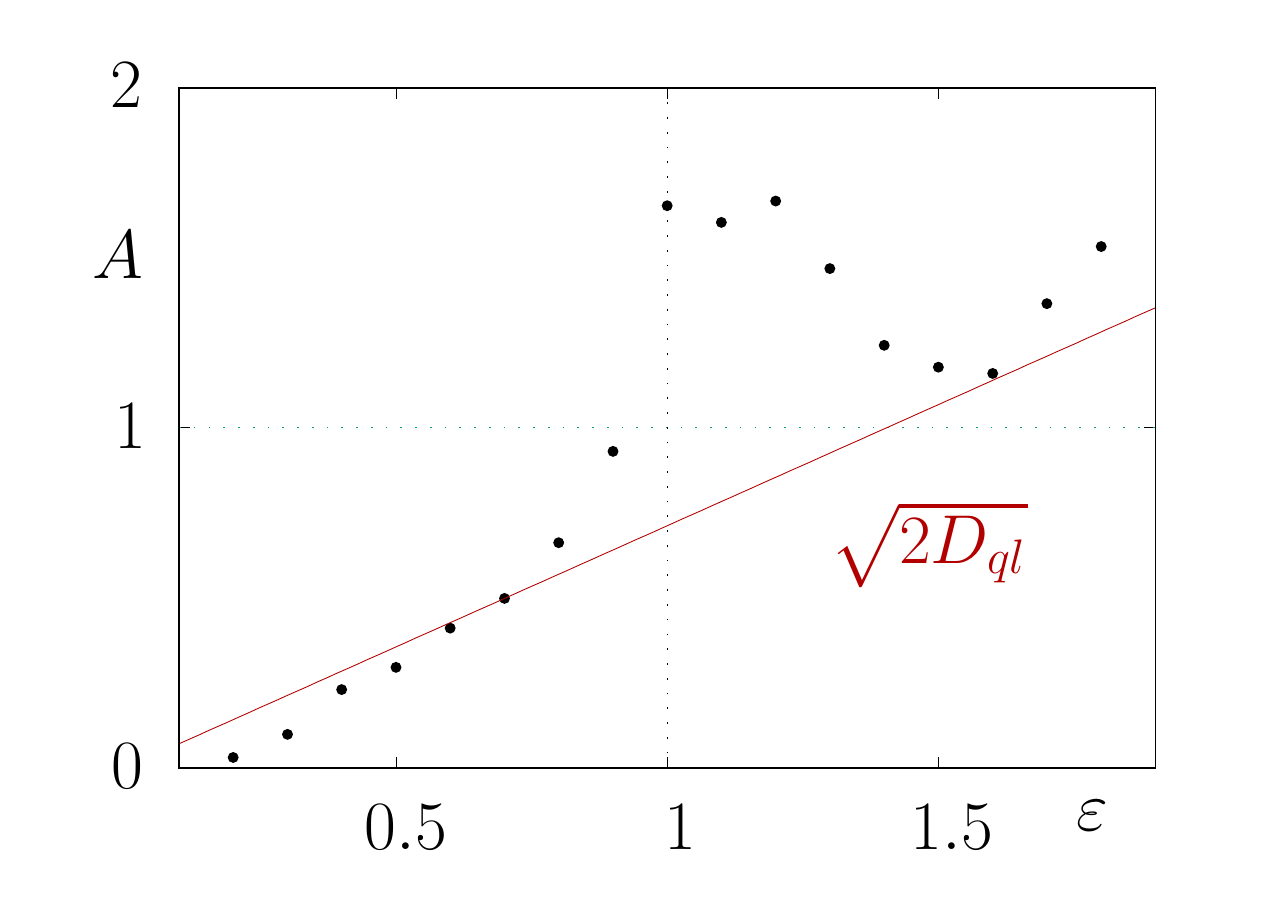}&
\hspace{-0.75cm}%
\includegraphics[width = 0.35\textwidth]{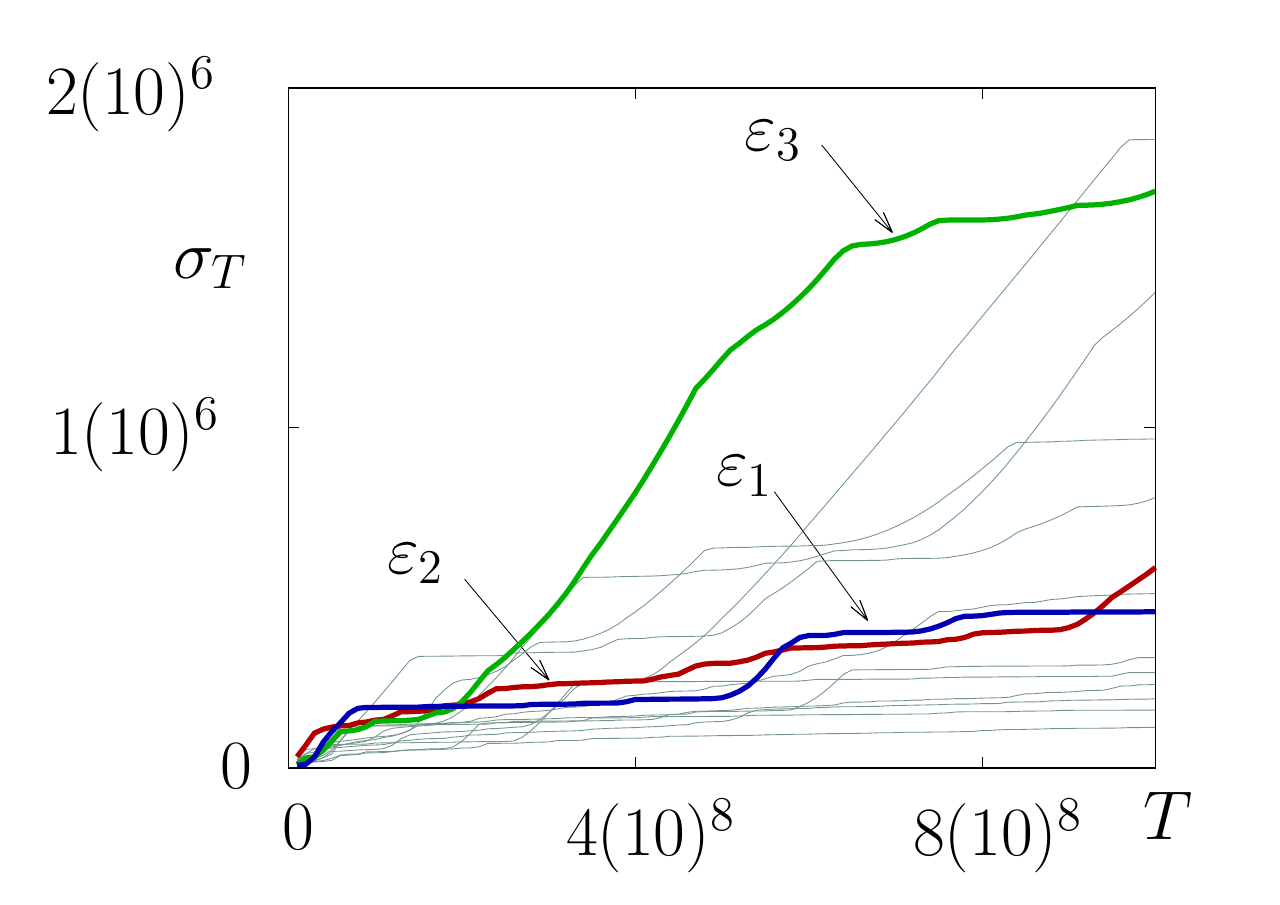}
\end{tabular}
\end{center}
\caption{
The standard deviation $\sigma_T$ as $\eps$ varies.  Left: The standard
deviation as a function of $T$ for nine values, $\eps = 0.2(0.1)1$, in black,
and eight values, $\eps = 1.1(0.1)1.8$, in red. Center: Growth rate $A$,
defined by $\sigma_T \sim A \sqrt{T}$ for these  $\eps$ values.  Right: The
standard deviation for fourteen (non-equispaced) $\eps\in[1.0005,1.0055]$. The
labelled curves correspond to $\eps_1=1.0007$ (blue), $\eps_2=1.0015$ (green)
and $\eps_3=1.0040$ (red).
}
\label{fig:std3D}
\end{figure}

The left panel of Fig.~\ref{fig:std3D} shows the standard deviation
\eqref{eq:sigmaT} as a function of $T$ for seventeen values of $\eps \in
[0.2,1.8]$. When $\eps < 1$ (black curves) there are no accelerator modes and
when $\eps = 1$, there are no bubbles. When $\eps \ge 1.1$ (red curves) the
accelerator bubbles have already disappeared. 

From this data it seems reasonable to assert that $\sigma_T \sim \sqrt{T}$. To
check this claim, we performed least squares fits of the full data sets for
each displayed $\eps$ to a function of the form $\sigma_T=AT^\chi$. For all
fits, we found $\chi\in(0.4975,0.5025)$, close to the expected value of
$\tfrac12$.  The corresponding values of $A$ are displayed in the central plot
of Fig.~\ref{fig:std3D} (black dots), together with the estimate
$\sqrt{2D_{ql}}$ (in red), recall \eqref{eq:quasilinear3D}. The deviation
between the numerically obtained values and the quasilinear prediction is
larger for $\eps \approx 1$ and the effect of the accelerator mode
can be seen even when there is no bubble. Note that when $\eps<1$ the diffusion
coefficient appears to grow nearly linearly with $\eps$, but at a slope
larger than the quasilinear estimate.
Recall that for Chirikov's standard map, the
quasilinear prediction is a better approximation for large parameter values
\cite{Chi79, RRW81, Ven08, MSV15}, but we have not checked values of $\eps$ larger than $1.8$ here.

In the right panel of Fig.~\ref{fig:std3D} we see that when
$\eps\in[1.0005,1.0055]$---when the FPAM around $P_\pm$ have stable
bubbles---$\sigma_T$ grows more rapidly than $\sqrt{T}$ and depends irregularly
on $\eps$.  Intervals of linear growth, corresponding to very long trapping
segments, are interspersed with intervals of slower growth where the orbit is not trapped or has only
short trapped intervals.  The considerable variability in the
growth of $\sigma_T$ as a function of $\eps$ is presumably due to the strong
dependence of the geometry of the bubbles on $\eps$ and to the sensitivity of
the long trapping times to chaos. 

\begin{figure}[ht]
\begin{center}
\begin{tabular}{ccc}
\includegraphics[width = 0.3\textwidth]{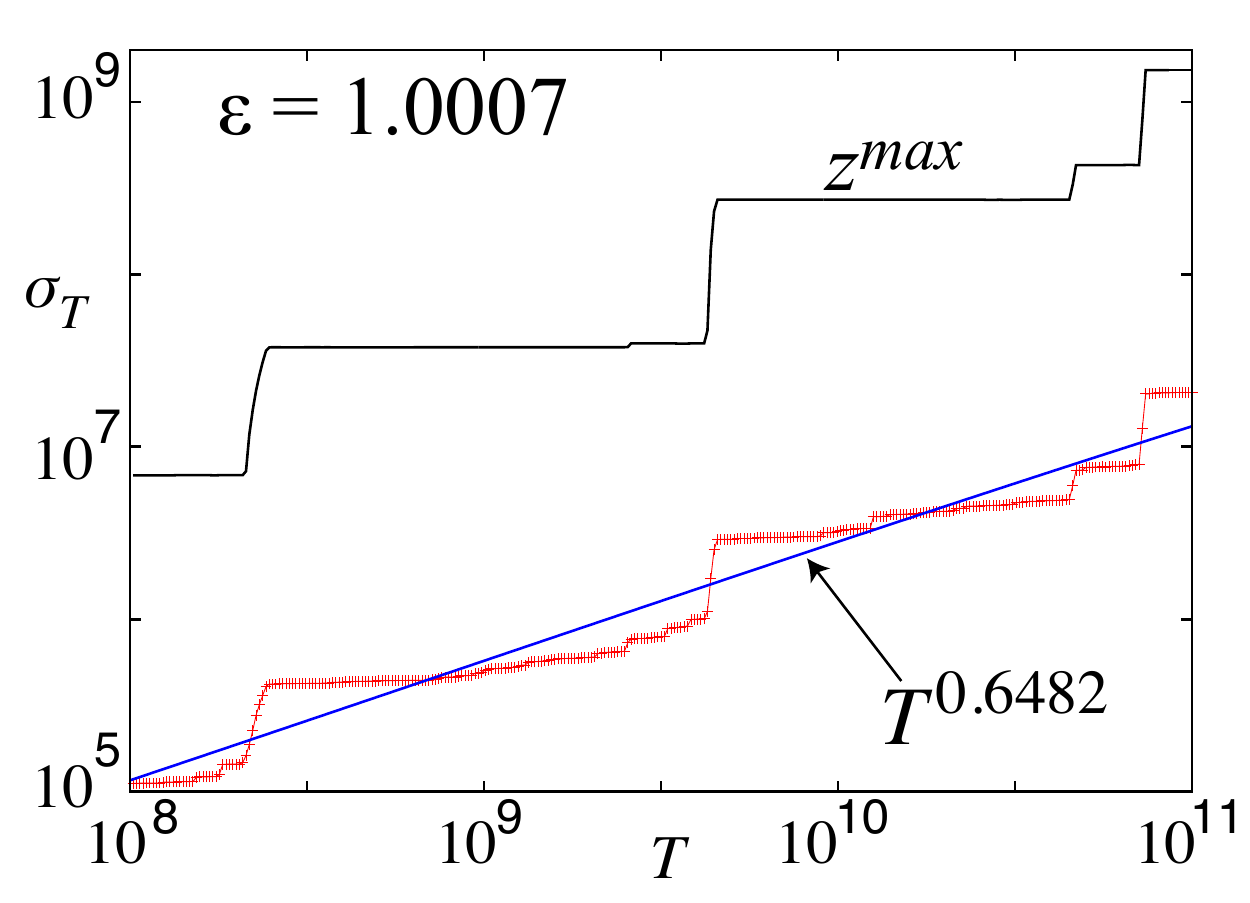}&
\includegraphics[width = 0.3\textwidth]{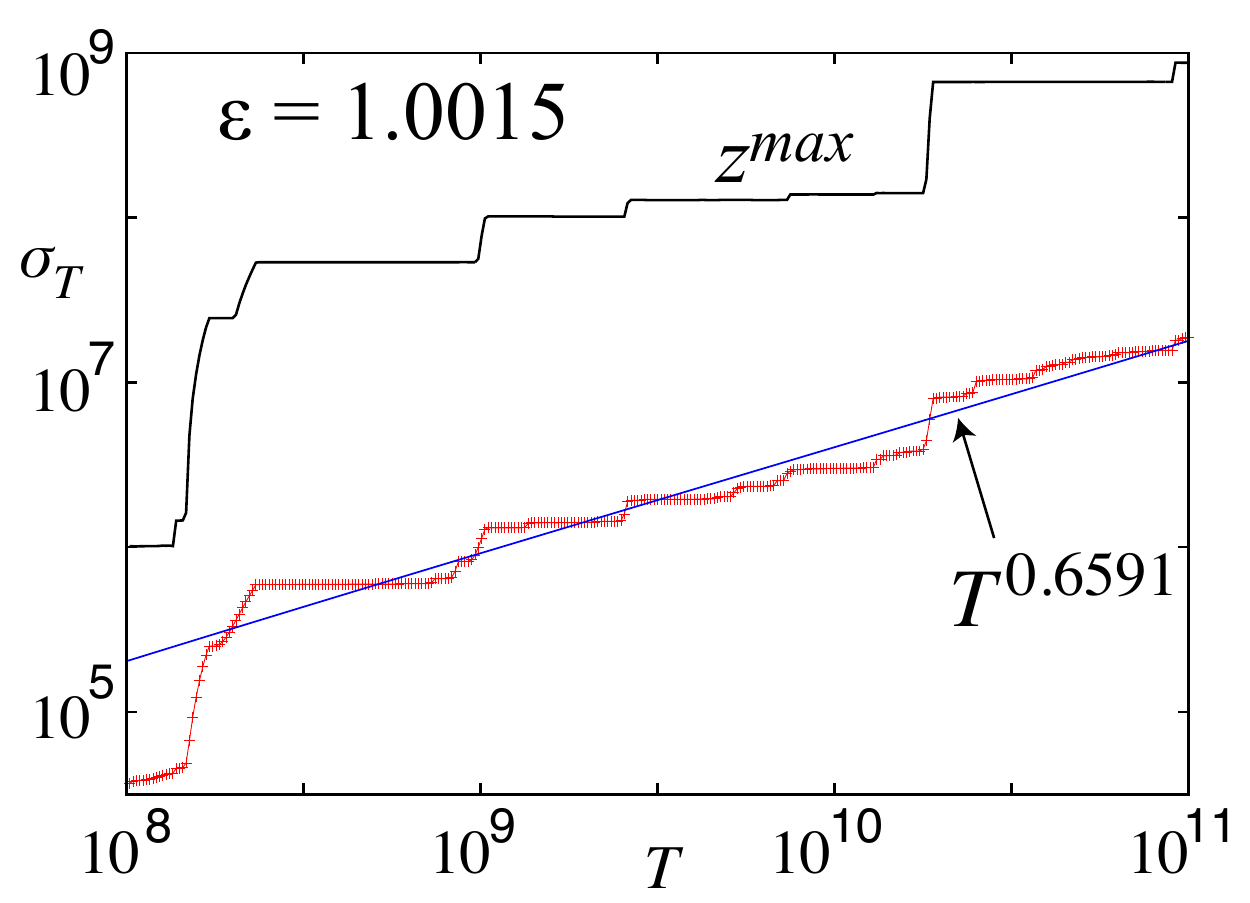}&
\includegraphics[width = 0.3\textwidth]{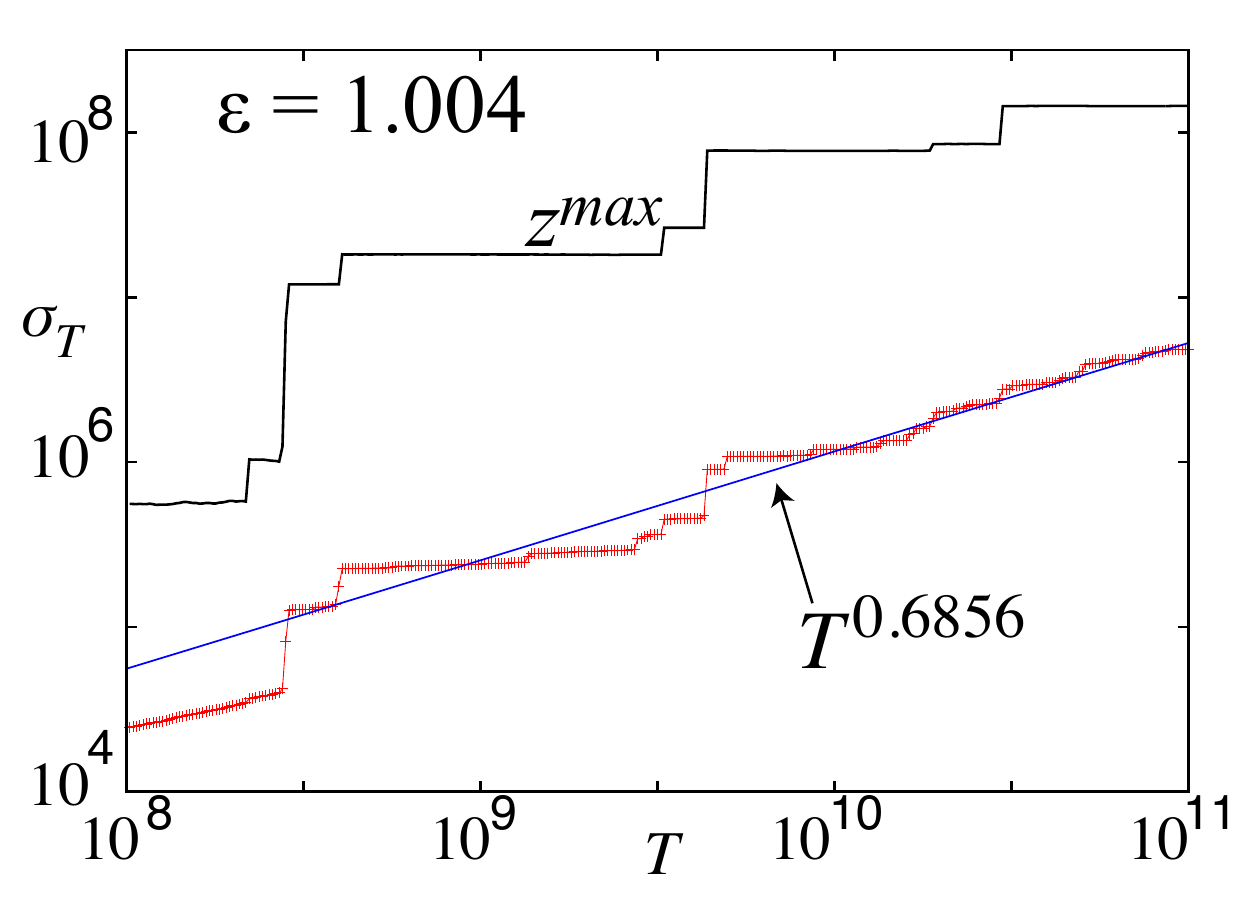}\\
\end{tabular}
\end{center} 
\caption{The standard deviation as a function of $T$ (red curves) on a log-log
scale for the $\eps$ values shown. A least squares linear fit (blue) gives the
slopes, $\chi$, indicated in each plot. The upper curves (black) show the
maximal value of $|z_T|$ among the $N=10^4$ initial conditions iterated.} 
\label{fig:anom3D}
\end{figure}

To assess the anomalous diffusive properties of $f_\eps$ we iterated $N=10^{4}$
initial conditions to $T=10^{11}$ to compute $\sigma_T$ for the three
particular values, $\eps_1$, $\eps_2$ and $\eps_3$---the highlighted values in
Fig.~\ref{fig:std3D}.  Logarithmic plots of the averaged $\sigma_T$ are shown
in Fig.~\ref{fig:anom3D}.  In these plots, a trapping interval can cause jumps
in
$$
	z^{max}_T = \max_{(x_0,y_0,z_0)} (|z_T|),
$$ 
sometimes up to an order of magnitude over a time interval of order $10^8$. In the previous definition
$(x_0,y_0,z_0)$ ranges in the set of initial conditions.

For the three $\eps$ values of Fig.~\ref{fig:anom3D}, a fit to $\sigma_T = A
T^\chi$ over $10^8 < t < 10^{11}$  gives exponents shown in Tbl.~\ref{tbl:Exponents}.
All are significantly larger than the diffusive value $\tfrac12$. Note that the value of $\chi$ depends
on the range of values used for $T$. In particular, it abruptly changes if we
end the simulation just before or after a big jump.

\begin{table}[tbp]
 \centering
 \begin{tabular}{c|ccc}      
    $\eps$   &$\chi$  & $b$    & $\chi+b/2$\\
    \hline
	$1.0007$ & 0.6482 & 2.0989 & 1.6977 \\
	$1.0015$ &  0.6591 & 2.4243 & 1.8713 \\
	$1.004\,\,\,$  &  0.6856 & 2.5630 & 1.9671 
\end{tabular}
   \caption{Exponent $\chi$ for the standard deviation \eqref{eq:sigmaT}, and $b$ for the exit time distribution
    \eqref{eq:trappingstat}-\eqref{eq:trappingprob} obtained from the numerical experiments on the map $f_\eps$ \eqref{eq:map3D} for the values $\eps_1$, $\eps_2$, and $\eps_3$. See \S\ref{subsect:remarks} concerning the last column.}
   \label{tbl:Exponents}
\end{table}

\subsubsection{Trapping statistics}\label{sect:algdecaytrap}

\begin{figure}[ht!]
\begin{center}
\begin{tabular}{m{8cm}m{8cm}}
\includegraphics[width = 0.45\textwidth]{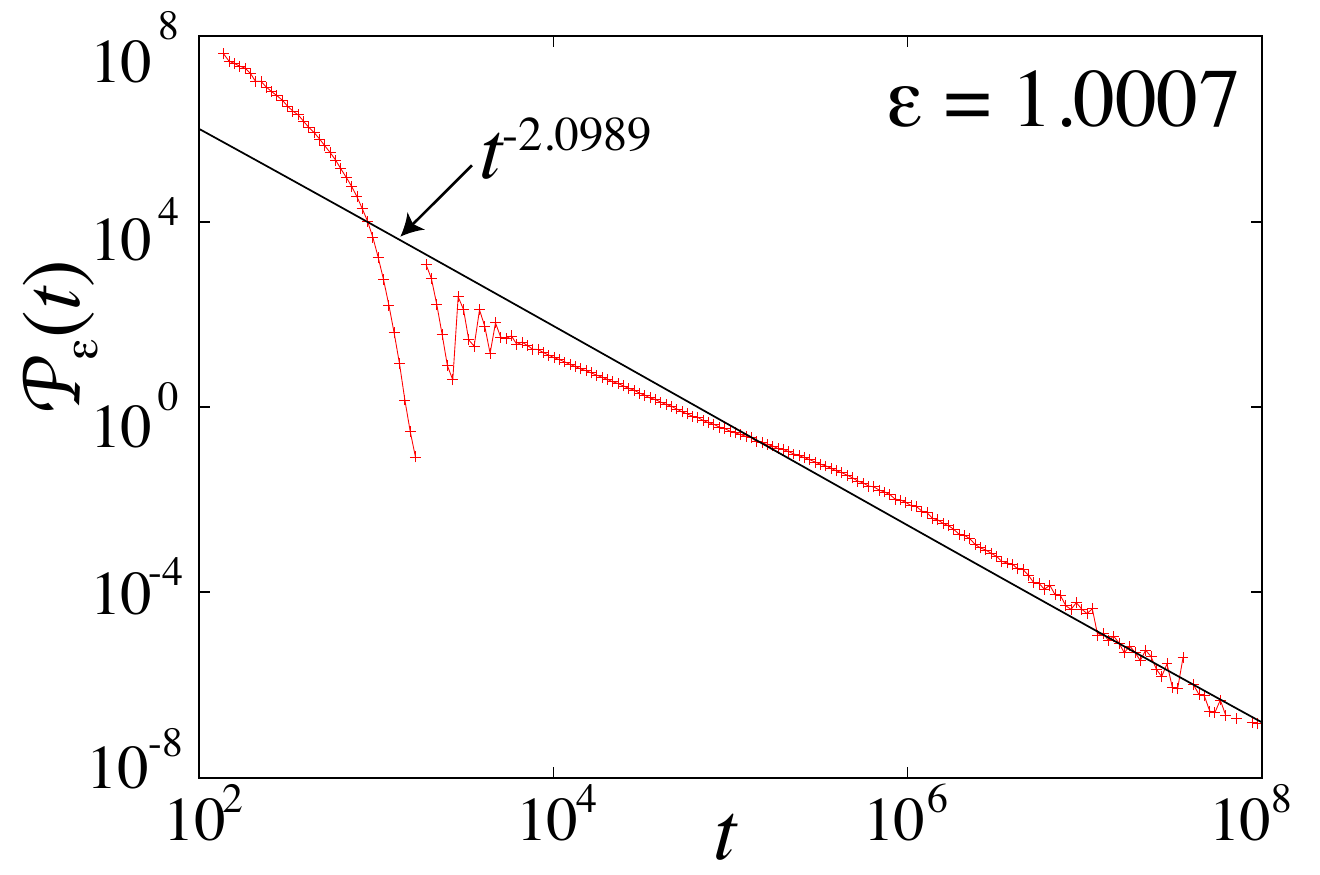}&
\includegraphics[width = 0.45\textwidth]{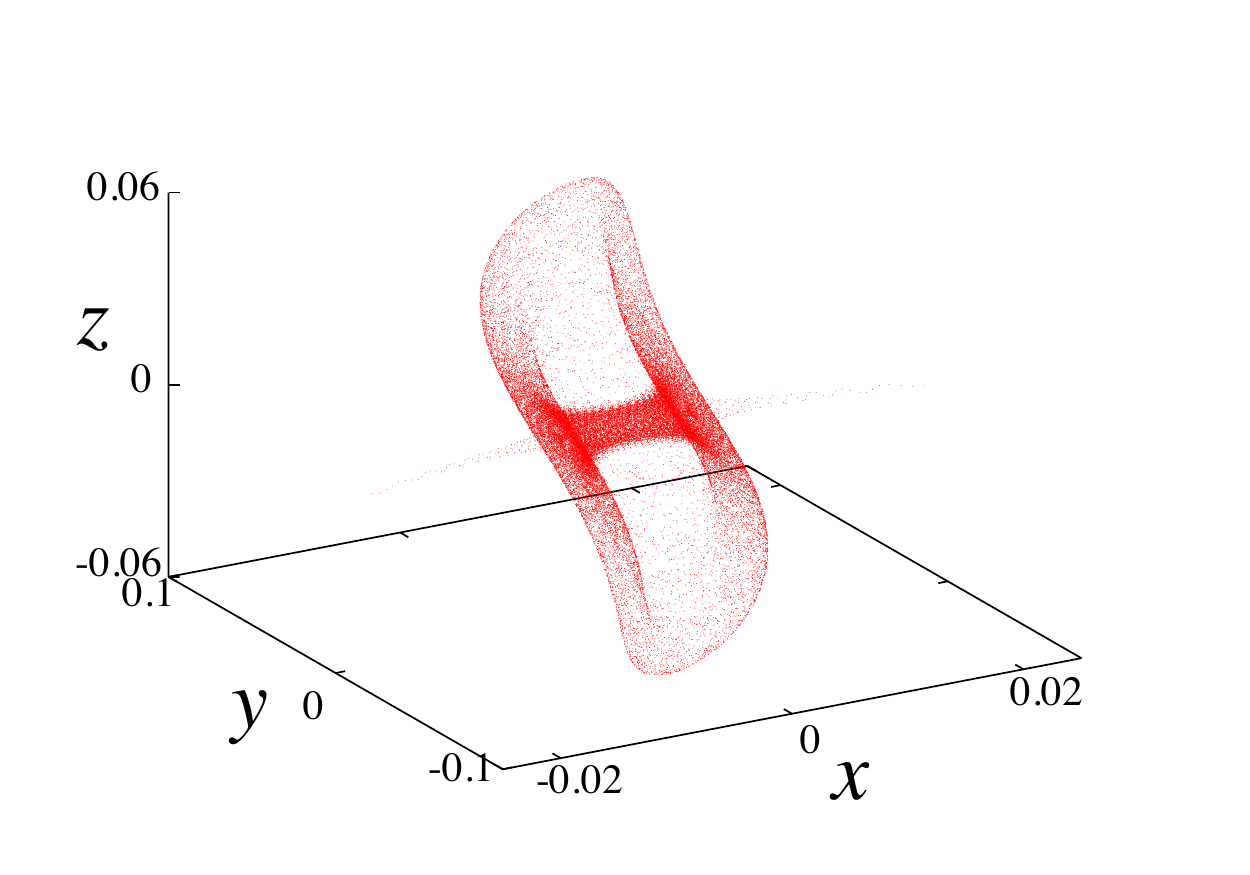}\\ 
\includegraphics[width = 0.45\textwidth]{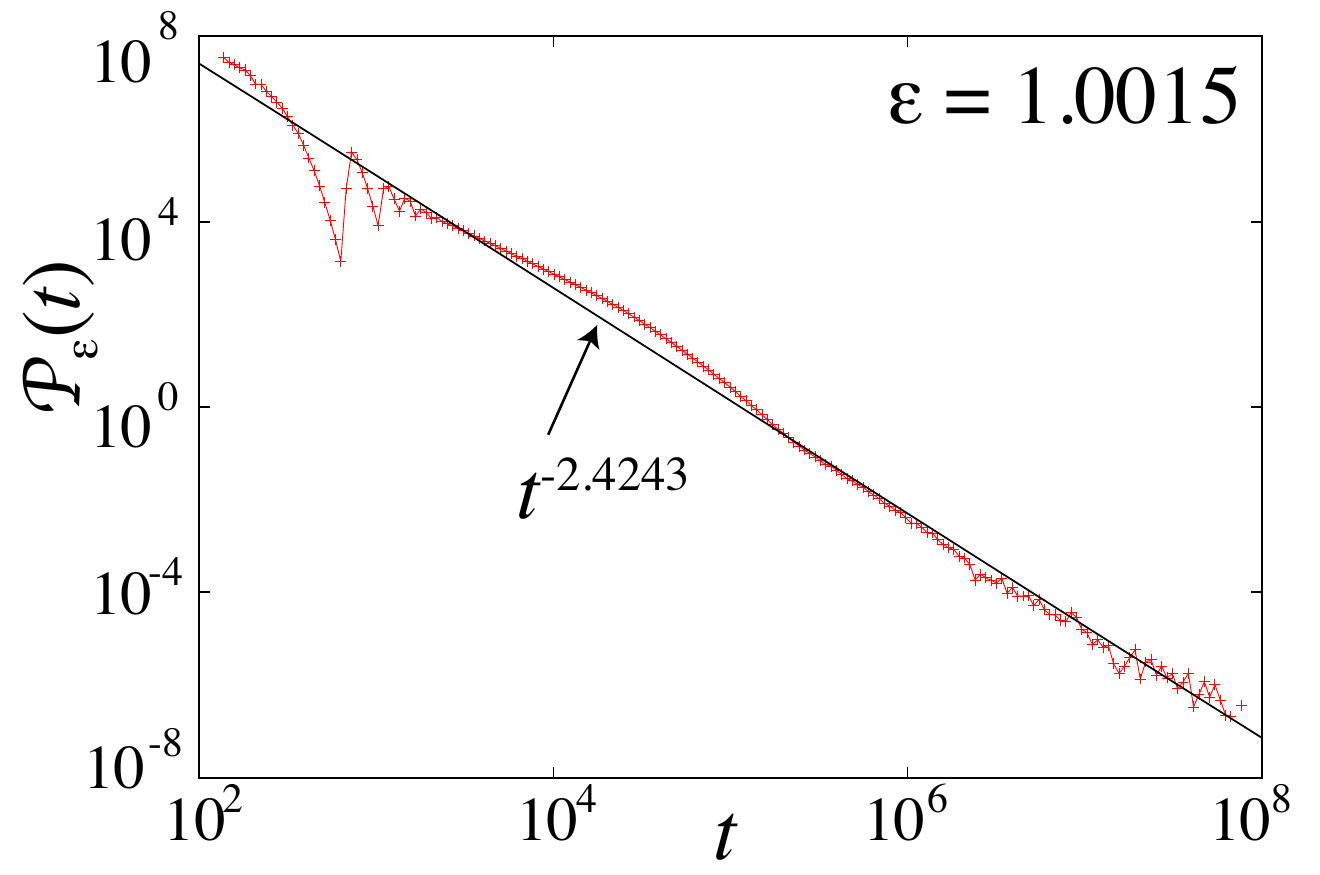}&
\includegraphics[width = 0.45\textwidth]{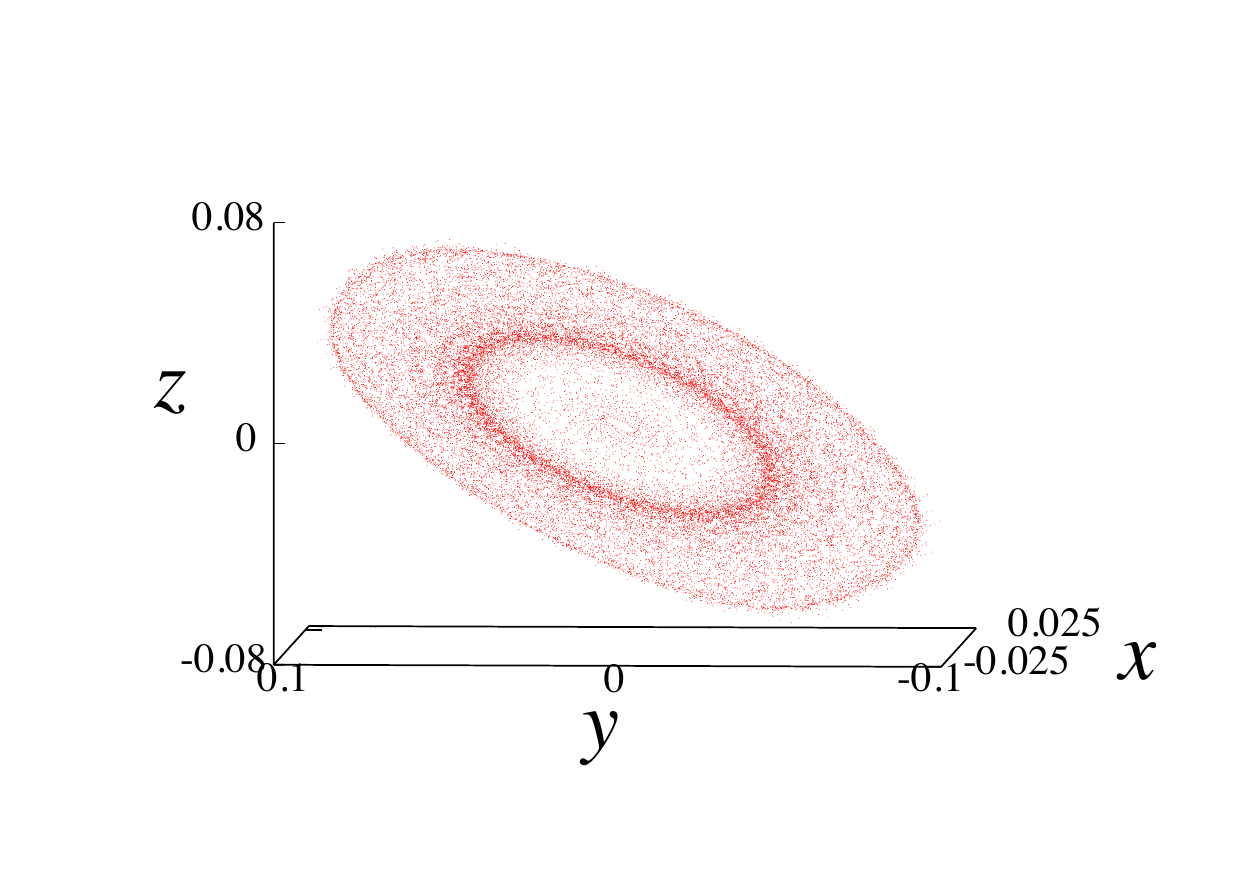}\\
\includegraphics[width = 0.45\textwidth]{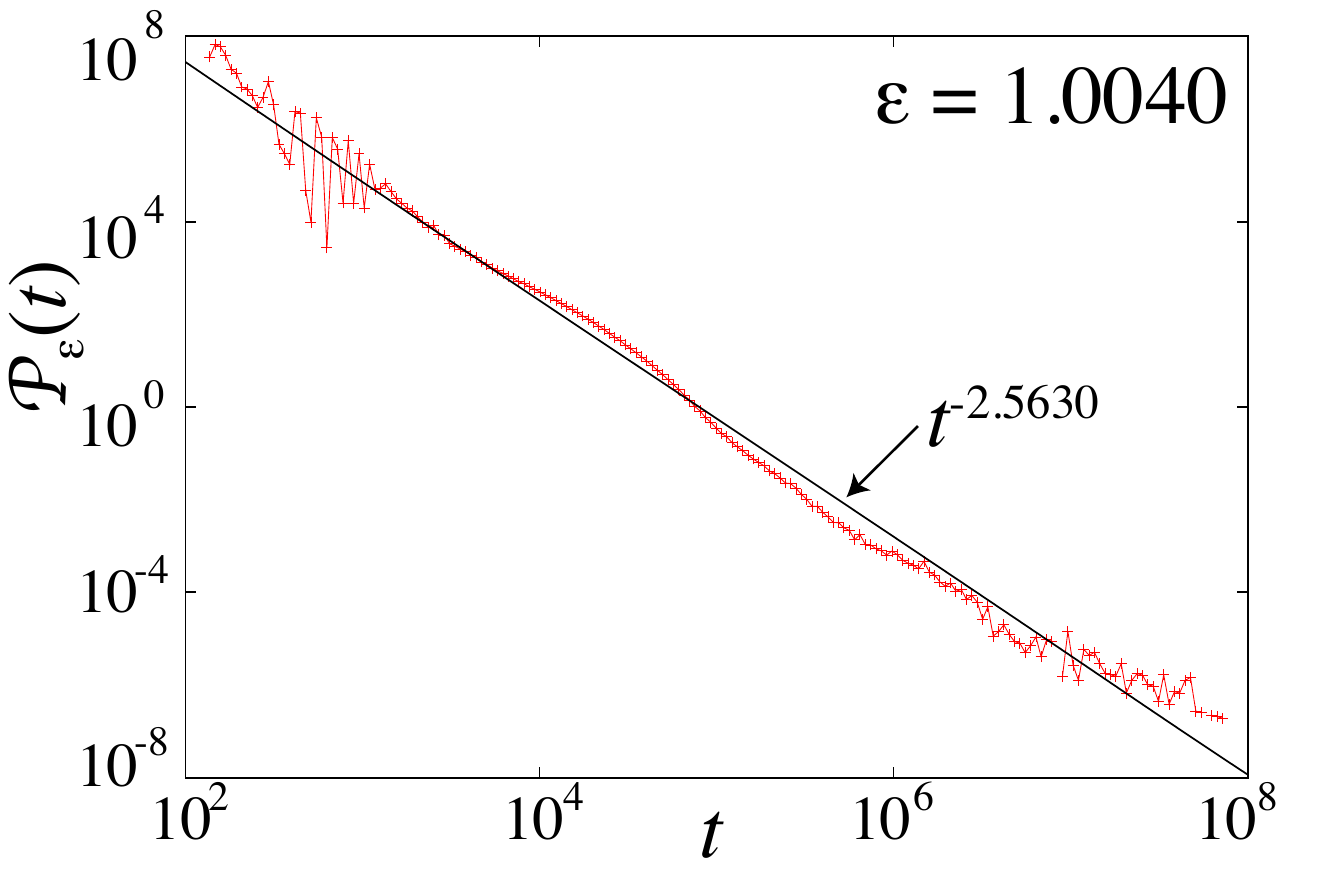}&
\includegraphics[width = 0.45\textwidth]{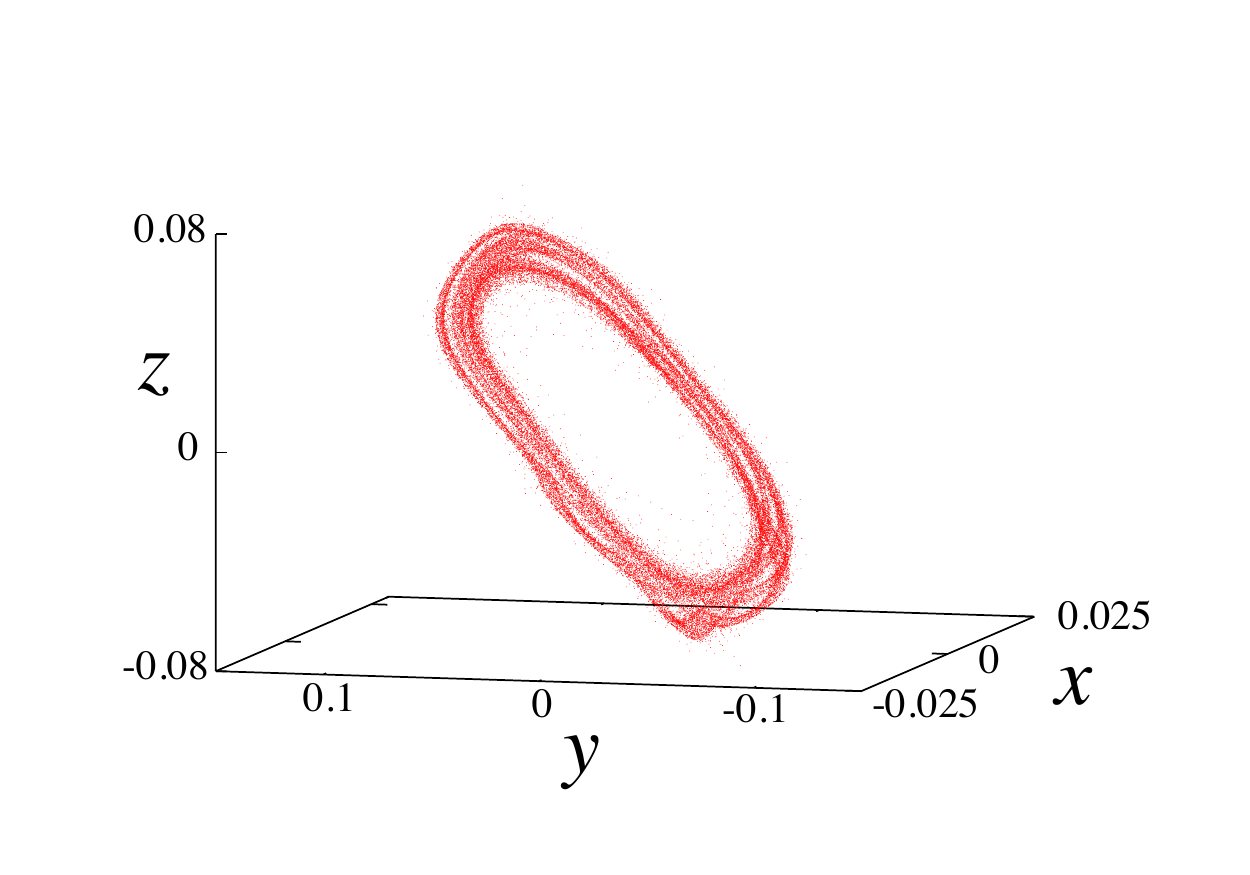}\\
\end{tabular}
\end{center}
\caption{Trapping statistics versus time for $\eps=1.0007,1.0015$ and $1.0040$.
Each right panel shows an example of a trapped orbit near the bubble of $P_+$
for the corresponding $\eps$ value on the left.}
\label{fig:trap3D}
\end{figure}

The trapping statistics \eqref{eq:trappingprob} for bubbles at $\eps_1,\eps_2$,
and $\eps_3$ are shown in log-log plots in  Fig.~\ref{fig:trap3D}. In all cases
it seems plausible to assume, following \eqref{eq:trappingstat}, that
$\Pcal_\eps(t)\sim t^{-b}$, with some fluctuations. A least-squares fit
(performed over the entire range) to a straight line (black) gives the
exponents shown in Tbl~\ref{tbl:Exponents}.
Repeating the computations for $\eps_1$ with $N=10^{6}$ initial
conditions and $10^{10}$ iterates gives the same value of $b$ to three decimal
figures. Such a power law decay was previously observed for a volume-preserving
map in \cite{MJM08}; by contrast in \cite{SZ09} the authors observe an
exponential decay of trapping statistics for another type of map.

\begin{figure}[ht]
\begin{center}
\begin{tabular}{ccc}
\includegraphics[width = 0.30\textwidth]{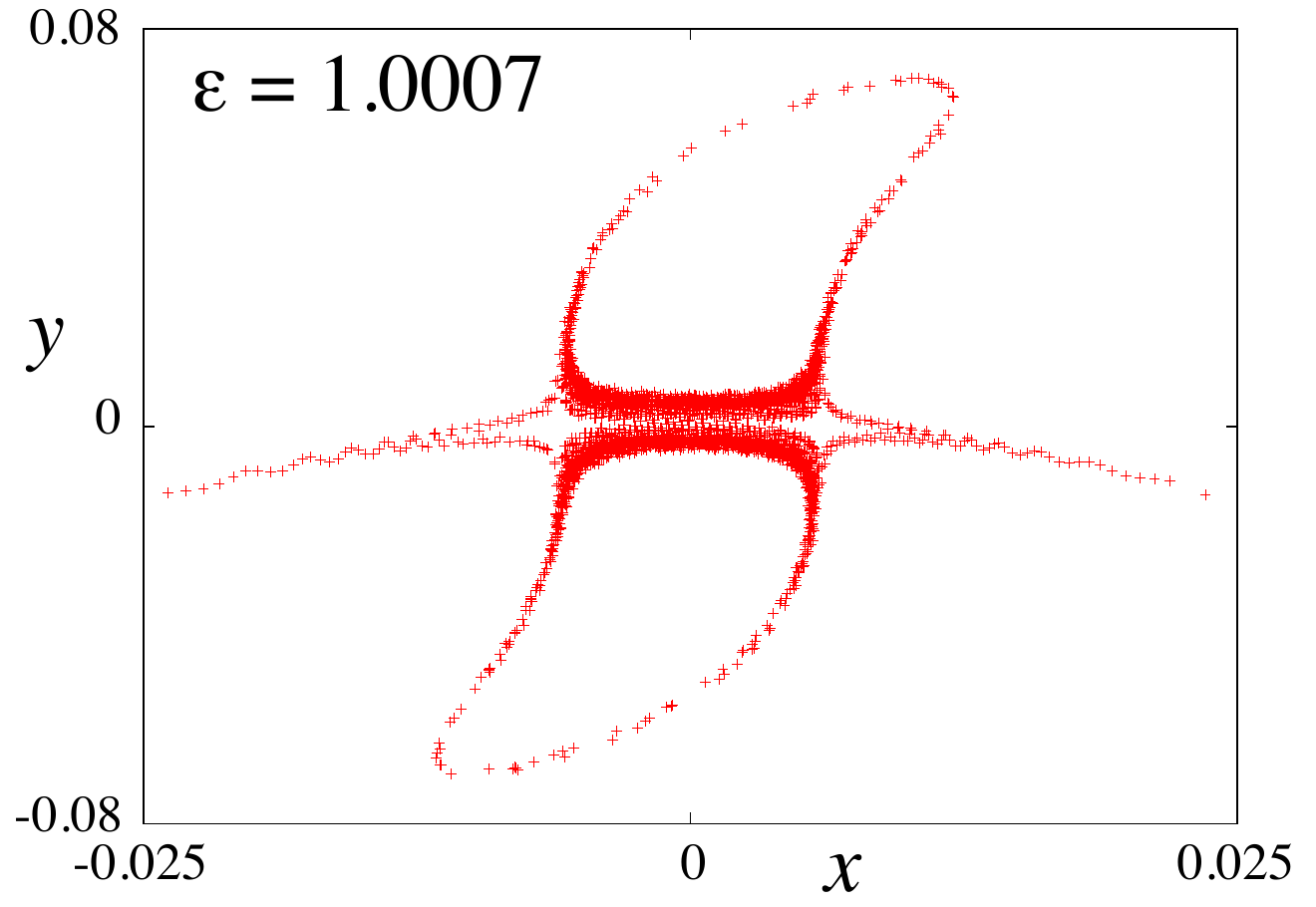}&
\includegraphics[width = 0.30\textwidth]{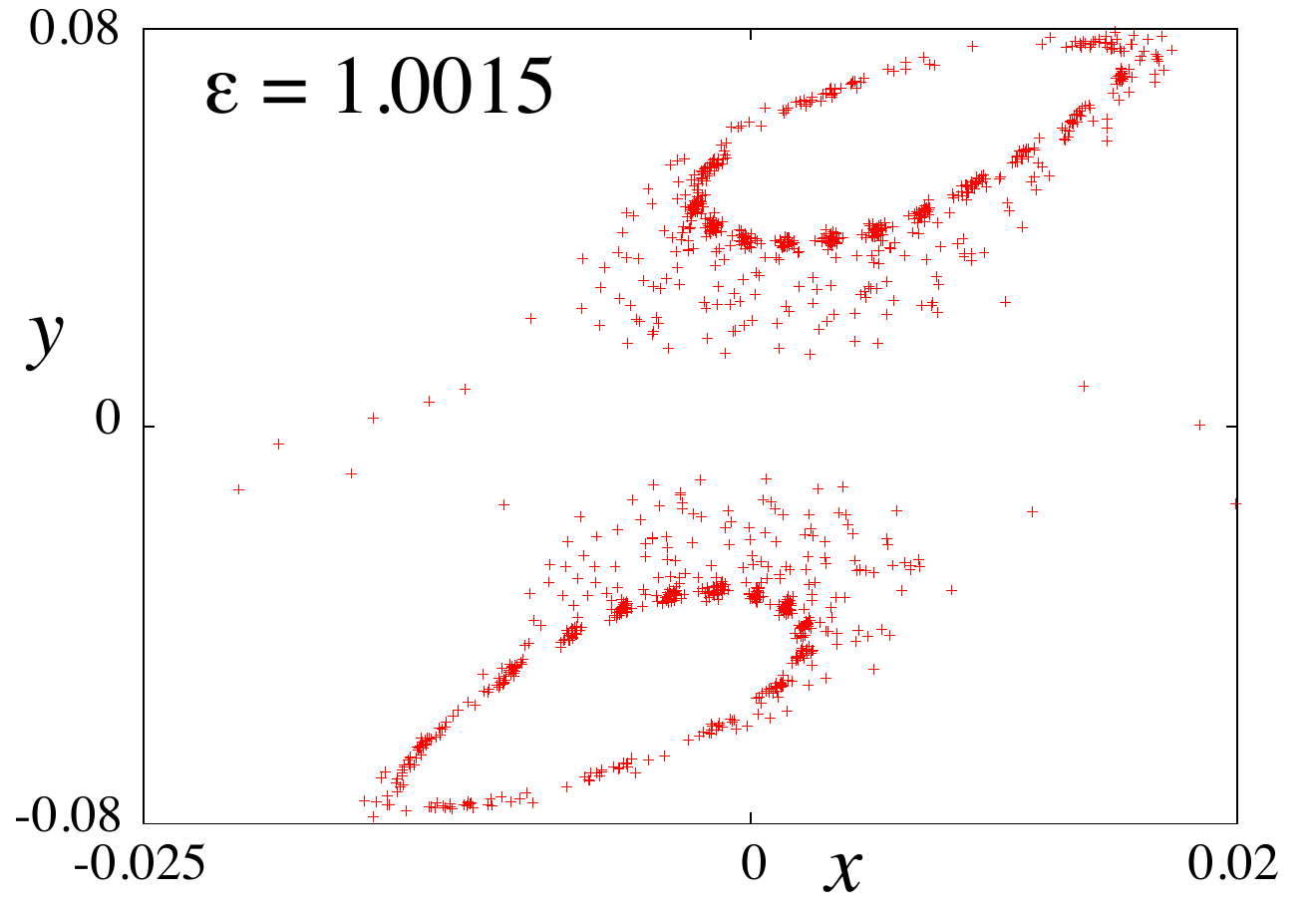}& 
\includegraphics[width = 0.30\textwidth]{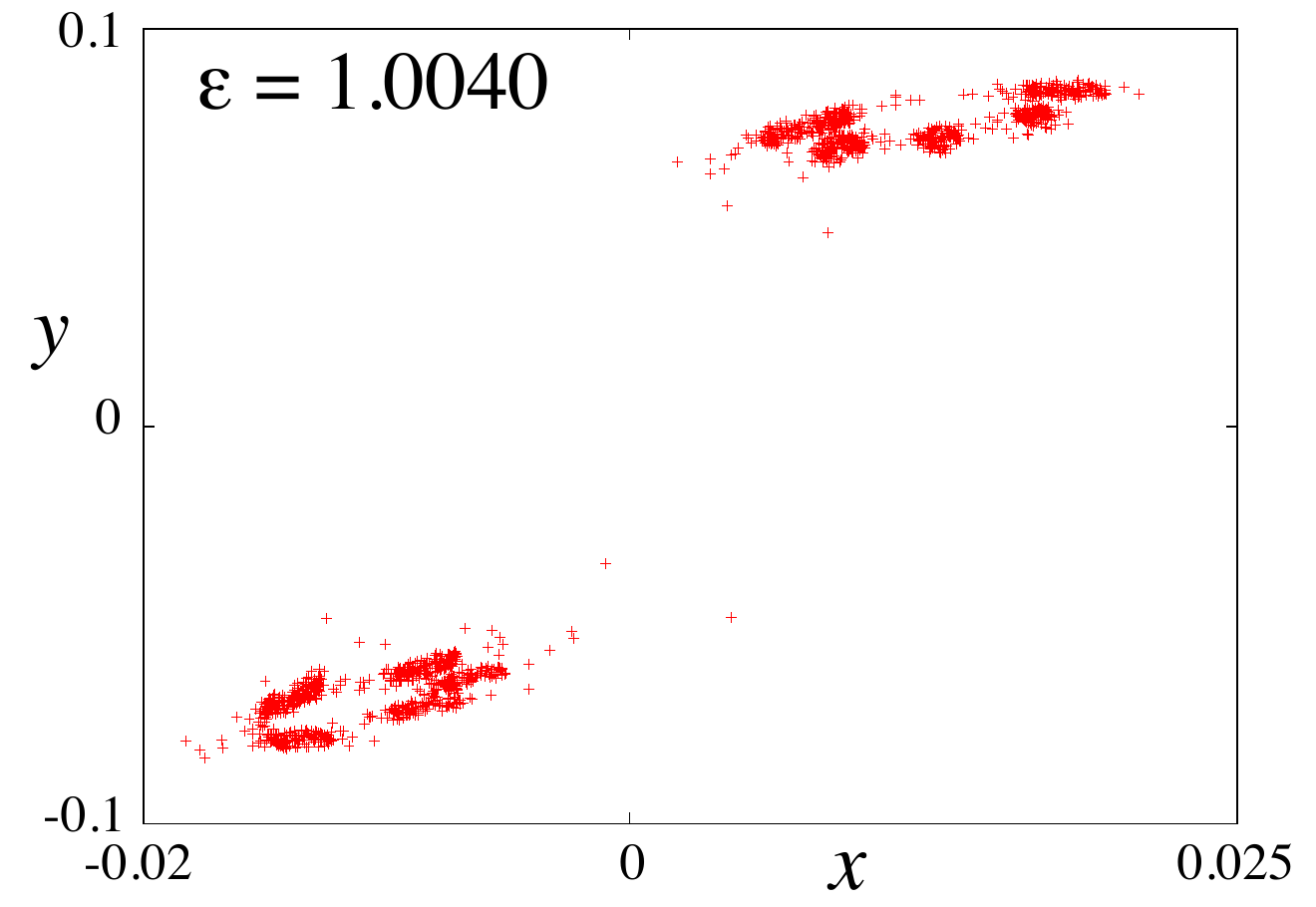}\\
\includegraphics[width = 0.30\textwidth]{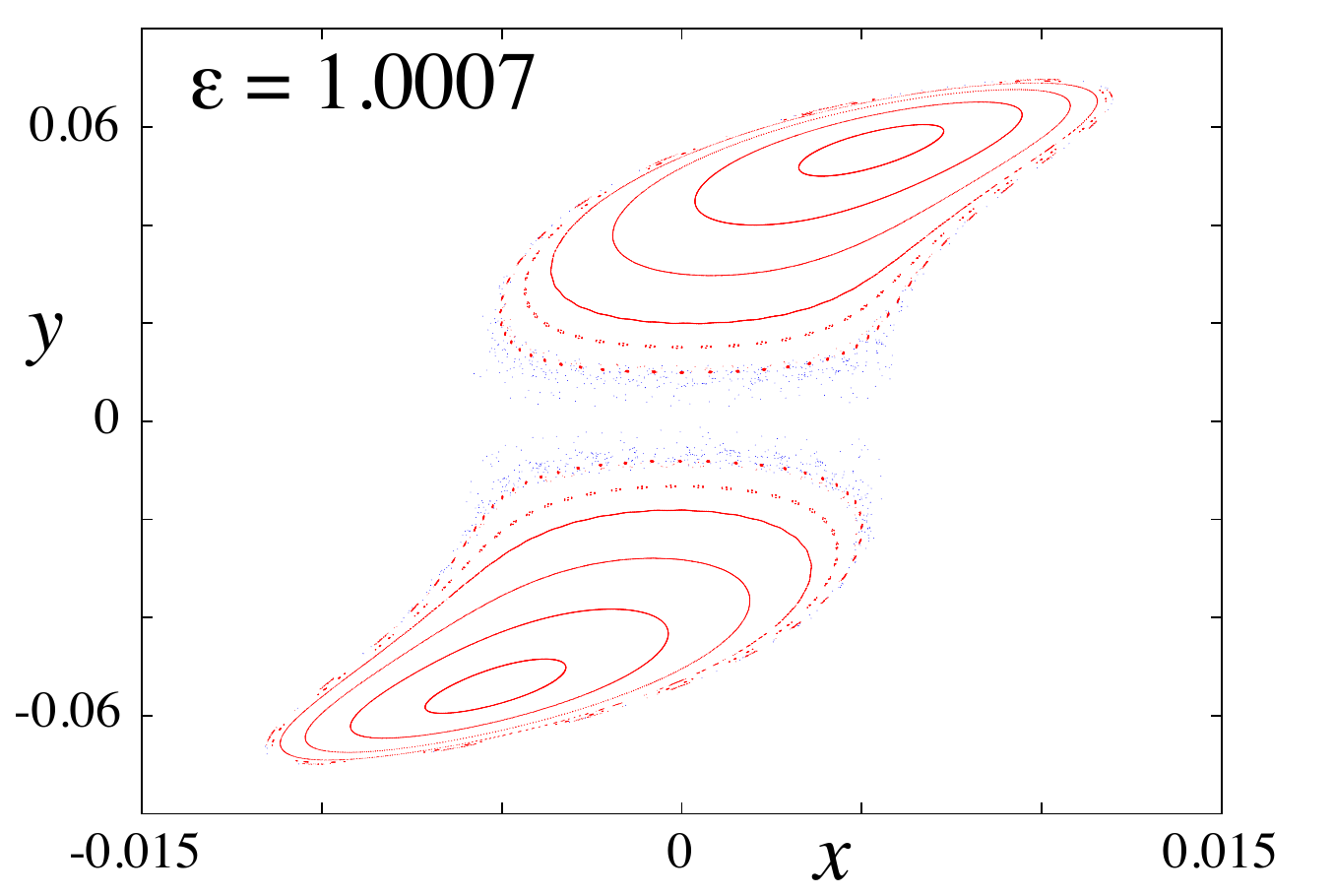}&
\includegraphics[width = 0.30\textwidth]{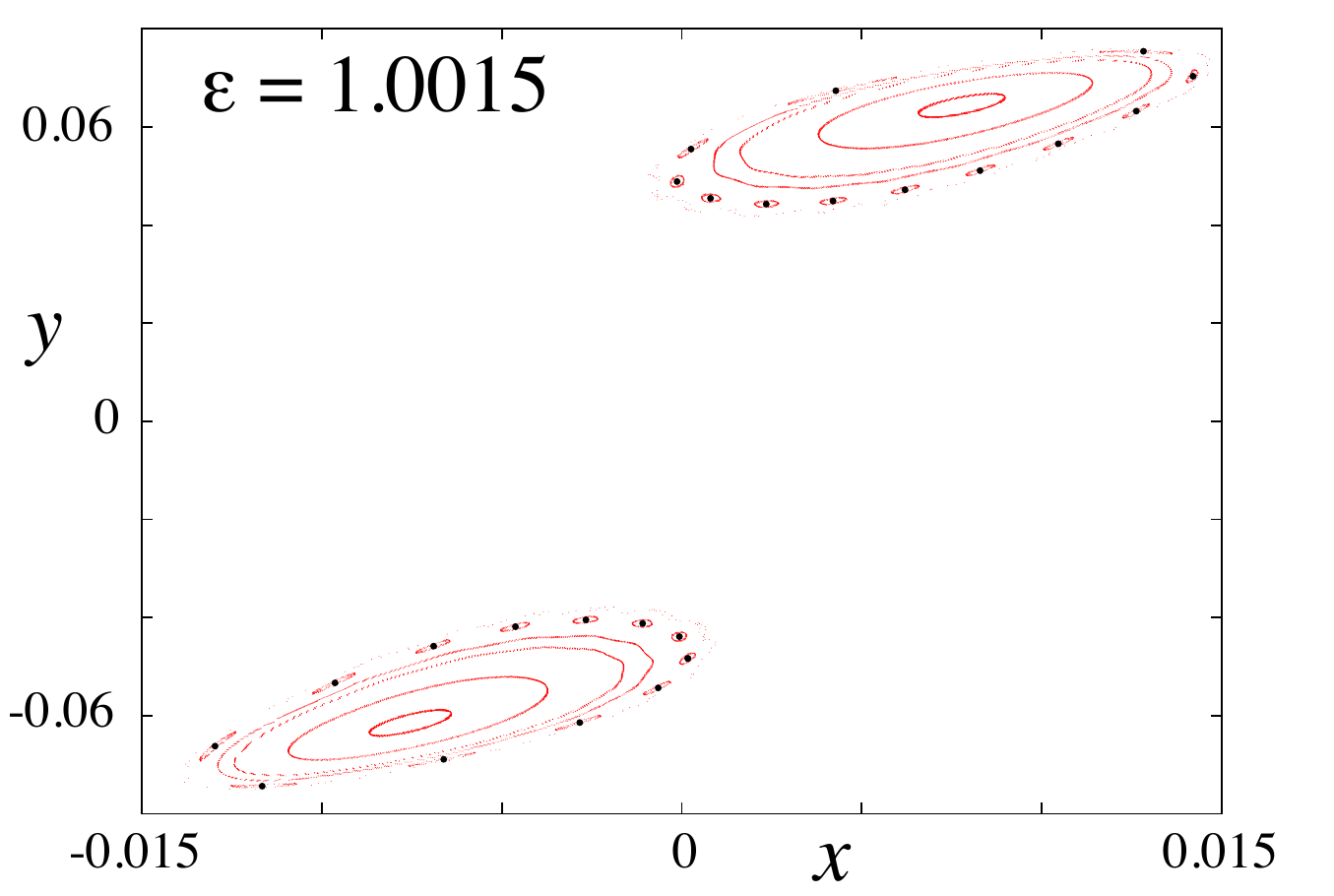}& 
\includegraphics[width = 0.30\textwidth]{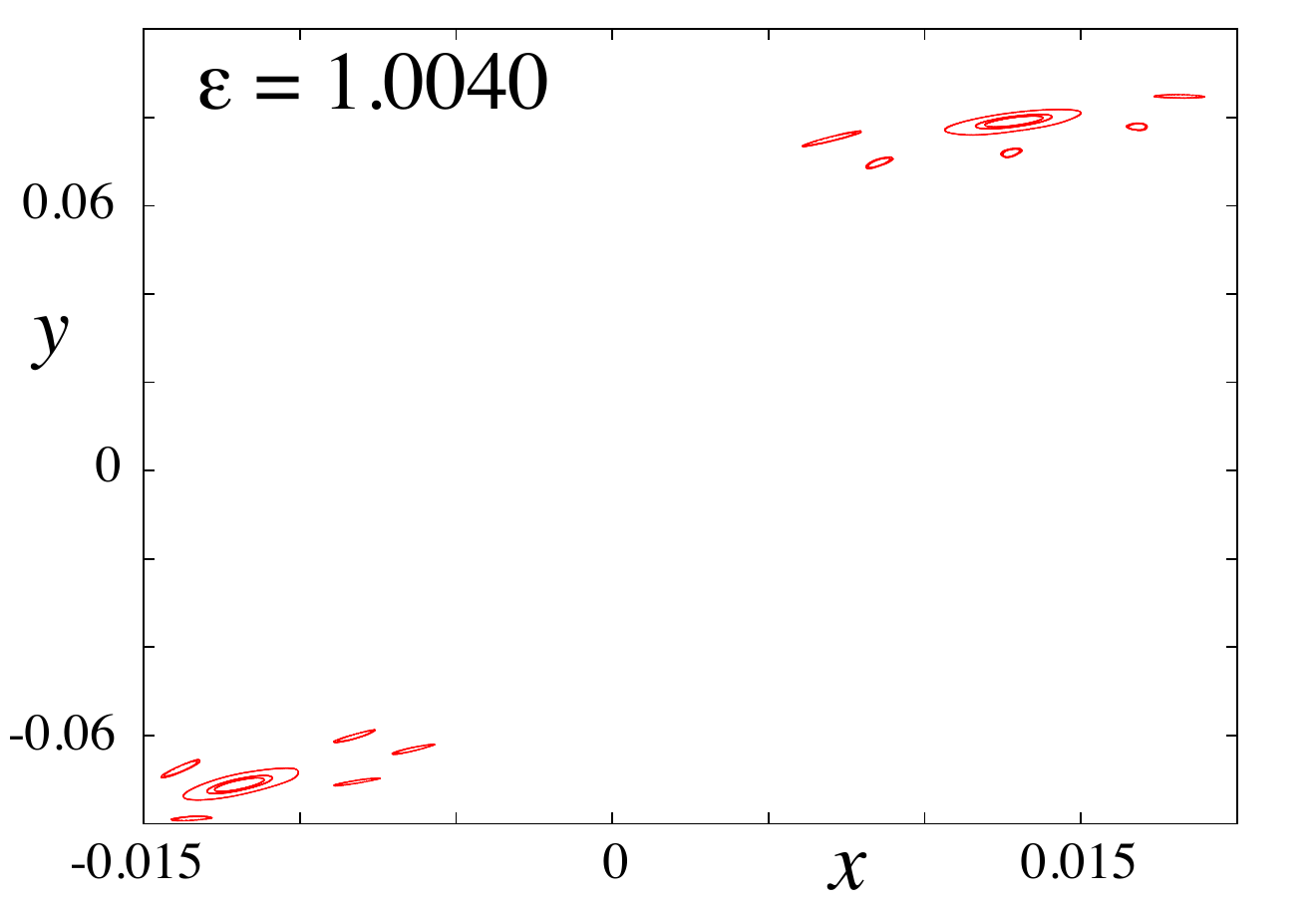}\\
\end{tabular}
\end{center} 
\caption{Slices through orbits near the bubbles of Fig.~\ref{fig:trap3D}.
Points on the orbits with $|z|<r$ are shown projected onto the $(x,y)$ plane.
Top row: Temporarily trapped orbits for $r = 0.01$ (left and middle) and $r =
0.02$ (right).  Bottom row: Trapped orbits inside the bubbles (red,
$r=10^{-4}$) and, for $\eps = 1.0007$, an escaping orbit (blue, $r=10^{-4}$).
For the middle plot the value of $r=10^{-7}$ has been used for the black
points.} 
\label{fig:trap3D-slice}
\end{figure}

Each panel in the right column of Fig.~\ref{fig:trap3D} shows a typical orbit
trapped near $P_+$ for the same $\eps$ as the left column.  Slices  near $z=0$
of these same orbits are shown in the $(x,y)$ plane in the top row of
Fig.~\ref{fig:trap3D-slice}. The bottom row of this figure shows slices through
some regular orbits in the $P_+$ bubble. Recall that when $\kappa_1>0$ the
point $P_+$ bifurcates into a pair of accelerating orbits
$P_+^{l,r}=(x^{l,r},0,0)$ \eqref{eq:xlr}. 

In \S\ref{sect:hopfone} we noted that
$P_+^r$ ($P_+^l$) has a 1D stable (unstable) invariant manifold and a 2D
unstable (stable) manifold.  These seem to play an important role in the trapping, and we will discuss
this in \S\ref{subsect:geom}.

For our three standard values of the parameter, we observe the following. 
\begin{itemize}
	\item{\bf $\eps_1=1.0007$}. Close to the birth of the bubble (recall
Fig.~\ref{fig:bounded_actualAM}) the invariant manifolds of $P^{l,r}_+$ can
be clearly guessed in Fig.~\ref{fig:trap3D-slice}. The longest trapped orbits
approach the bubble along $W^s(P_+^r)$, then follows a trajectory that
seems to cover a 2D torus, finally escaping along $W^u(P_+^l)$.
	\item {\bf $\eps_2=1.0015$}.  Further away from the birth of the bubble
there are prominent satellite tori outside the main tori, and the longest
trapped orbits appear to be primarily stuck around such satellites: in
Fig.~\ref{fig:trap3D-slice} this region has the highest density. Each of these
satellites encloses an elliptic invariant circle giving what seems to be a
period-twelve orbit in the section (the black points in the bottom middle
panel of Fig.~\ref{fig:trap3D-slice}). In fact, there are six invariant curves
of $f_\eps^6$, one the image of the other under $f_\eps$. Under $f_\eps^6$ each
of these curves closes after two revolutions around the $x$ axis. The central
region of the bubble, near the 1D manifolds of $P^{l,r}_+$, has a lower
density, but it still seems to play a role in its stickiness.
	\item {\bf $\eps_3=1.0040$}. Now the regular region around the bubble
is almost destroyed, but one still expects trapping around the main tori or
satellite tori. The orbit shown in Fig.~\ref{fig:trap3D-slice} seems to be
trapped around a family of tori that surrounds a single elliptic
invariant curve, which closes after five revolutions around the $x$ axis. 
\end{itemize}

Recall that by Prop.~\ref{prop:dynloc},  the Michelson map
\eqref{eq:Michelson_map} is a quadratic approximation near $P_\pm$ for the
family  \eqref{eq:map3D}. Though this approximation is less accurate when
$n=1$, there is a coordinate change of the form $(x,y) \mapsto
(x-G(\eps)y^3,y)$, for suitable $G(\eps)$, that brings the plots in
Fig.~\ref{fig:trap3D-slice} closer to those in Fig.~\ref{fig:Michmap-slice} for
the Michelson map.


\section{Discussion}\label{sect:discussion3D}

In this section we discuss in more detail how chaotic orbits approach the vicinity of a
bubble. We also discuss how
the results of the previous section fit with, and deviate from, existing theoretical approaches, 
suggesting a possible approach to deal with the discrepancies. 

\subsection{Entering and exiting the bubbles} \label{subsect:geom}

\begin{figure}[h!]
\begin{center}
\begin{tabular}{cc}
	\includegraphics[width = 0.45\textwidth]{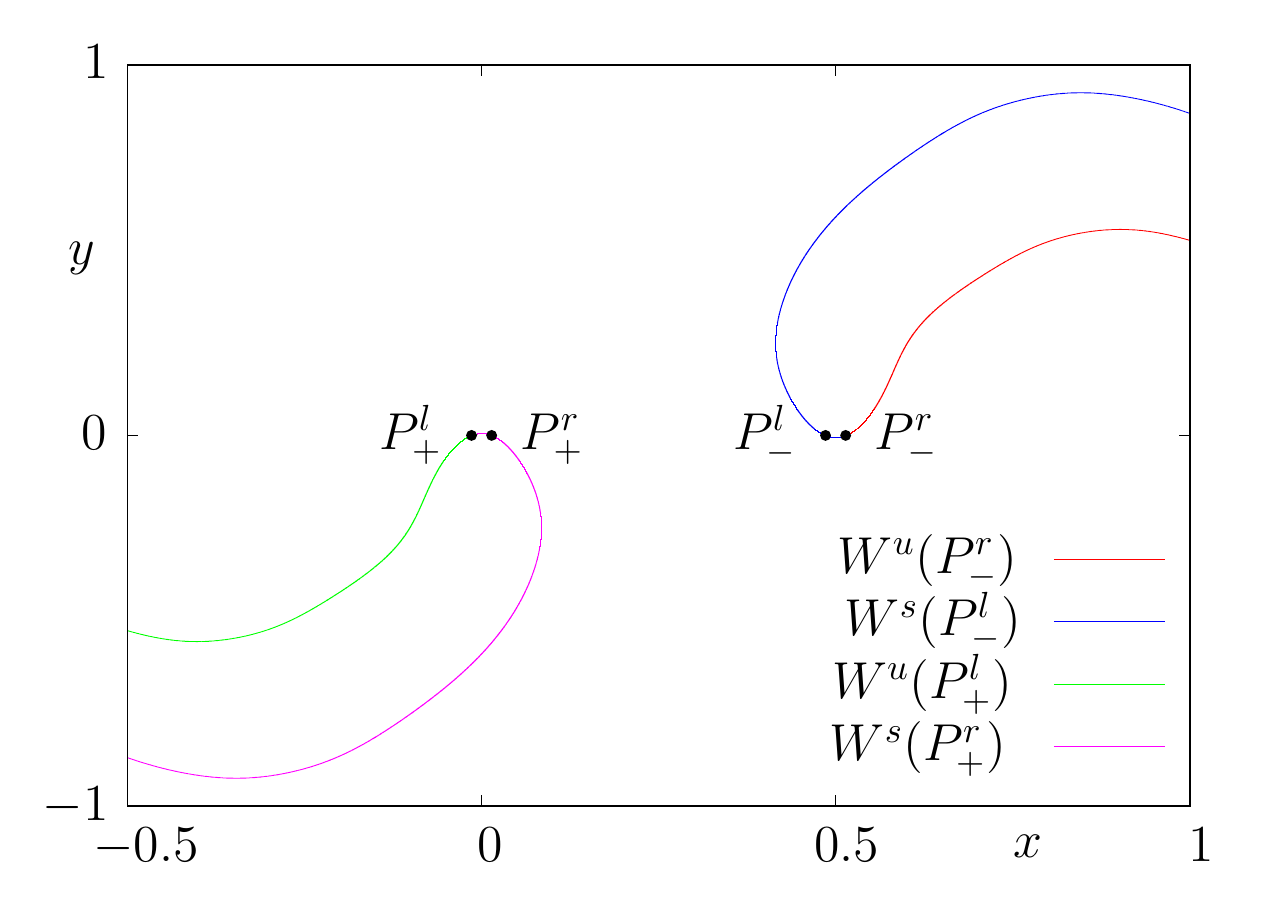}&
	\includegraphics[width = 0.45\textwidth]{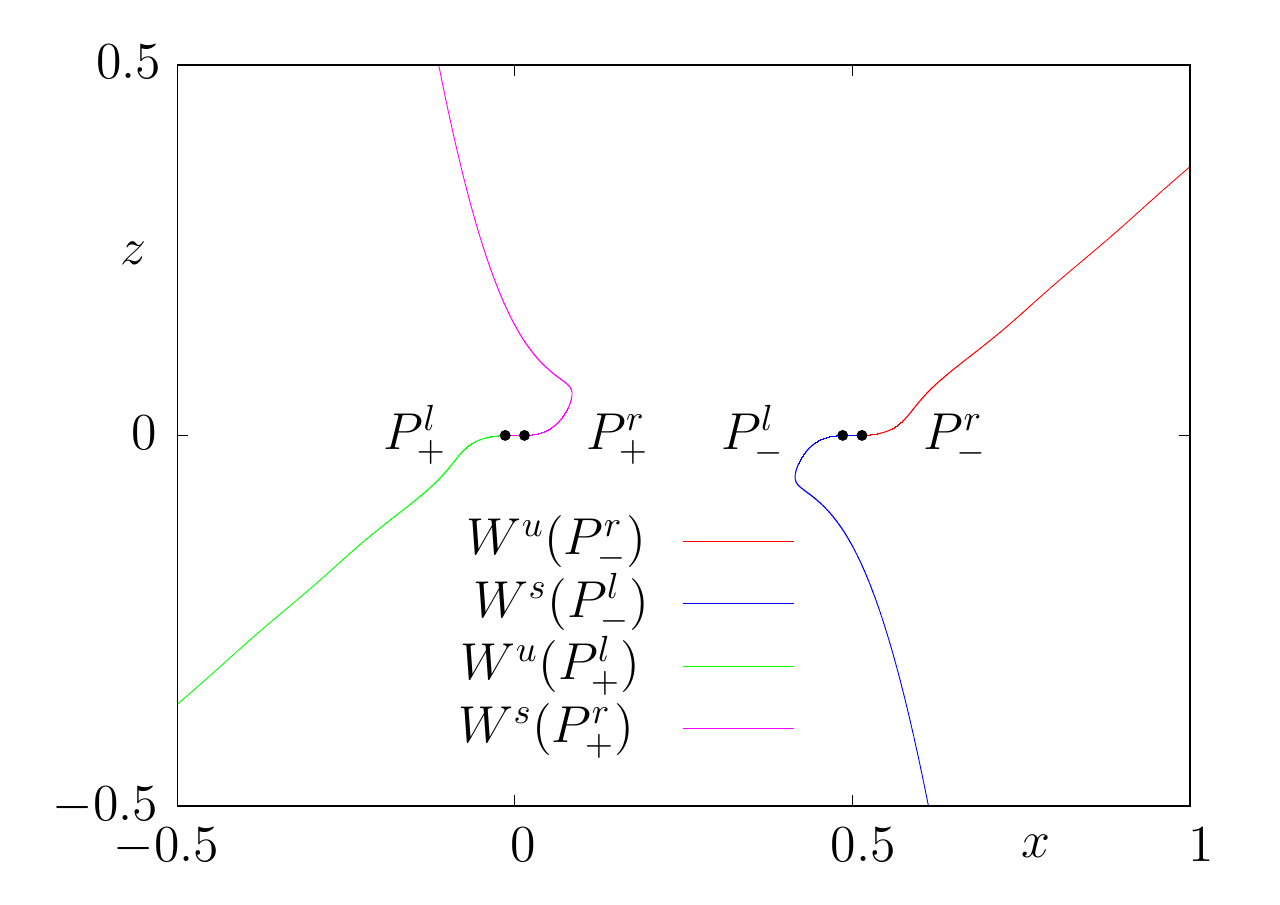}
\end{tabular}
\end{center}
\caption{One-dimensional manifolds of $P_{+,-}^{r,l}$ for $\eps_3=1.004$
projected into the $(x,y)$ (left) and $(x,z)$ (right) planes, shown for $x \in
[-0.5,1]$.}
\label{fig:invman}
\end{figure}

As we noted above, the entrance and exit routes for
a bubble often correspond to the 1D manifolds of the fixed points $P_{\pm}^{r,l}$ of $\tilde{f}_\eps$.
Numerical computations of these manifolds are shown in Fig.~\ref{fig:invman} for $\eps_3$. 
Qualitatively similar curves are obtained for other parameters. Recall that the reversing symmetry
\eqref{eq:reversor} implies that the invariant manifolds of $P_{-}^{r,l}$ can
be obtained from those of $P_{+}^{r,l}$ using the reversor \eqref{eq:reversor}, and this symmetry is clearly 
manifest in the figure.  When a bubble is present, points on outer branches of the unstable 1D manifolds
do not appear to return to a neighborhood of the bubbles in a short number of iterations. The implication is that
these manifolds correspond to entrance and exit routes for the neighborhood of a bubble.

A large fraction of orbits that get trapped in $\Wcal_{+}$ \eqref{eq:stickyregion}
approach $P_+^r$ along the right branch of $W^s(P_+^r)$, the purple curve in Fig.~\ref{fig:invman}. 
They then move away from this point along its 2D unstable manifold, $W^u(P_+^r)$ (not shown in the figure).
This manifold curves towards the neighboring saddle-focus, $P_+^l$. The 2D stable manifold of this point similarly curves towards $P_+^r$,
and so these two manifolds intersect. Some orbits are thus funneled along $W^s(P_+^l)$ towards $P_+^l$.
They finally escape the bubble close to the left branch of $W^u(P_+^l)$, the green curve in the figure. 
Though the incoming orbits to $\Wcal_{+}$ need not
be very close to $W^s(P_{+}^r)$, the attraction of $W^s(P_+^l)$ tends to make escaping orbits closely follow
$W^u(P_{-}^r)$. Moreover, the length of the trapped segment is longer if an 
orbit is closer to the stable manifolds, since such orbits spend more time near the saddle-foci. 
By symmetry the same explanation applies to incoming and escaping orbits for the region $\Wcal_{-}$
around $P_-^{l,r}$. The case $\eps = 1.0007$ in Fig.~\ref{fig:trap3D} and its corresponding slice around $z=0$ in
Fig.~\ref{fig:trap3D-slice} illustrate this situation.
 
If an orbit remains trapped for a long time, it will often follow a trajectory close to a boundary torus of 
the bubble (an outermost 2D torus). 
When such an orbit reaches the vicinity of $P_+^l$, it can be swept through the center of the bubble along 
the right branch of $W^u(P_+^l)$. This will lead to a return near $P_+^r$, and the orbit can repeat the process. 
A small number of trajectories make many turns inside the
bubble becoming trapped for a long time near sticky, 2D tori.
Each turn requires a passage close to the two saddle-foci where the orbit spends a relatively large
number of iterates.
The effect of repeated returns can be clearly seen in the trapping statistics plots of
Fig.~\ref{fig:trap3D} especially for $\eps_1$. Let us give some details on what is observed: 
\begin{enumerate}
\item First, orbits that enter the bubble and leave it without
being swept through the center, can escape more rapidly
from $\Wcal_{+}$ than those orbits that return close to $P_+^r$. This
creates a discontinuity in the trapping statistics. The same thing happens
for orbits that have multiple passages through the channel created by the 1D
manifolds: for each additional passage there is a new discontinuity. 
Consequently, the trapping statistics in the figure show corresponding jumps (for, say,
$10^3 \lessapprox t \lessapprox 10^4$ for $\eps_1$).

\item Second, the relative measure of orbits that do not perform any close return 
to $P_+^r$ decreases as the distance to the saddle-foci decreases. 
The implication is that there are more
orbits spending shorter times near the bubble than longer times.
For the statistics at $\eps_1$, this explains the decrease in the abundance of trapped orbits 
for, say, $10^2 \lessapprox t \lessapprox 10^3$. 
Similar effects are seen, but to a smaller extent, for the orbits that pass multiple times through
the channel. These effects are weaker, but still visible in the plots for
$\eps_2$ and $\eps_3$.
\end{enumerate}

As $\eps$ grows, the channel around the 1D manifolds that traps orbits grows in diameter, but can still play some
role. For example, the slices for $\eps = 1.0015$ in
Fig.~\ref{fig:trap3D-slice} show that some trapped orbits still can be stuck in
a zone with larger volume near the 1D manifolds. Of course if $\eps$ is large
enough this channel will be less important.

\subsection{A transport model} \label{subsect:stat}

A statistical model of transport usually assumes that ensembles evolve as a random
walk on a discrete Markov chain with states corresponding to regions of phase
space bounded by partial barriers.  For area-preserving maps, the barriers are
Cantori, and the transition flux between states is the turnstile
area~\cite{MMP84, Mei92, Mei15}.

A simplified model for trapping statistics and anomalous diffusion
corresponds to discretization into two such states \cite{AK08, IHKM91, WS96, ZK93, ZK94}: a region
$\Wcal = \Wcal_+\cup\Wcal_-$, \eqref{eq:stickyregion}, where orbits are accelerated, and its complement,
$$
	\Wcal^c = \Tset^3\setminus \Wcal.
$$
The idea is that when an orbit is in $\Wcal$ it undergoes a \textit{flight}, where the action grows linearly in time, and while it is in $\Wcal^c$ it undergoes normal diffusion.
In this model there are just two possible transitions: escape from, or
entry into $\Wcal$, i.e. the transitions $\Wcal\to\Wcal^c$ and $\Wcal^c\to\Wcal$, respectively. 
If we take $\Wcal$ to be a vicinity of a
bubble of stability, then this simplification requires that we know
the exit-time probability $\Pcal_\eps(t)$ \eqref{eq:trappingprob}, the pdf
of a $\Wcal\to\Wcal^c$ transition at time $t$.  From our observations it seems 
plausible to assume that this has the
power law form \eqref{eq:trappingstat} with $b\in(2,3)$. This is
consistent with previous numerical results for a 3D map \cite{MJM08} and with the
observations for 2D maps, recall \S\ref{sect:apm}.  
Note that $b$ must be at least $2$ since, when a map is volume
preserving, Kac's theorem implies that the average exit time must exist \cite{Mei97}.
When $b<3$, the variance does not exist.

Of course the true distributions
in Fig.~\ref{fig:trap3D} are not
exactly power laws: there are jumps and oscillations. The
former is probably due to low flux through regions containing newly broken tori, and the
latter to the number of passages close to the saddle-foci $P_{\pm}^{r,l}$ \cite[Ch. 5]{Mig16}. 

\begin{figure}[ht!]
\begin{center}
	\includegraphics[width = 0.55\textwidth]{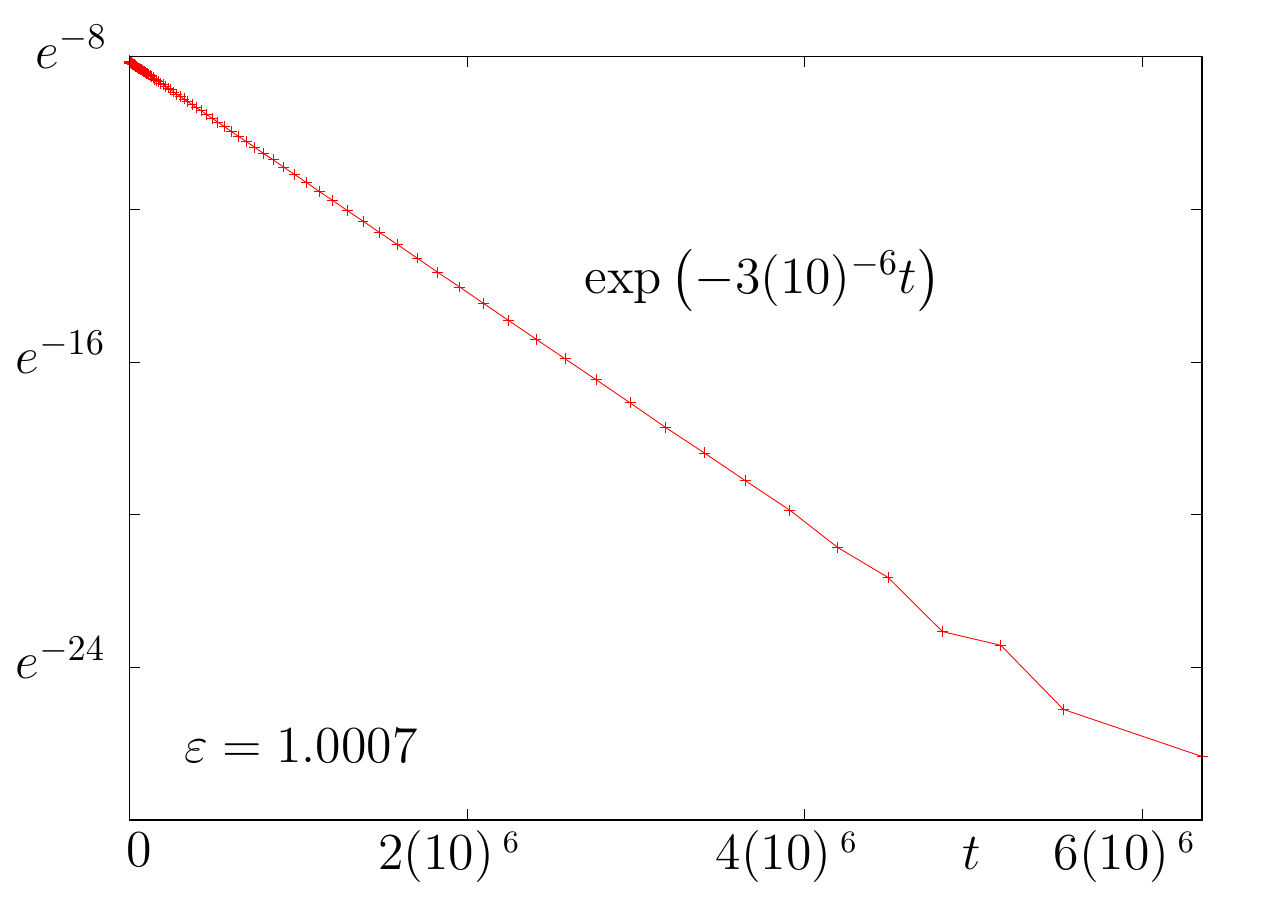}
\end{center}
\caption{Exit time probability density function for $\Wcal^c$ on a log-linear plot. The
distribution is computed for $N=2\times 10^7$ initial conditions in a fundamental domain of the
unstable manifold of the point $(3/4,0,0)$.}
\label{fig:zonaexp}
\end{figure}

The analogous pdf for the lengths of stays
outside $\Wcal$ is the exit-time distribution for $\Wcal^c$. 
As was also observed in the area-preserving context \cite{MSV15}, this 
distribution seems to be well approximated by an exponential. In
Fig.~\ref{fig:zonaexp} we show, for $\eps_1$, the exit time distribution
for $\Wcal^c$ as a function of time.  
In essence, excluding fast returns to $\Wcal$ (say, of length less than $50$), 
it appears that the probability of entering $\Wcal$ after
spending $t$ iterates in $\Wcal^c$ seems to follow a geometric distribution
with rate $c$, and hence that the exit time distribution for $\Wcal^c$ is
\begin{eqnarray}\label{eq:outtoin}
	\Pcal(t) \sim (1-c)^{t} \sim e^{-c t},
\end{eqnarray}
when $c \ll 1$.  Estimating $c$ from a linear fit on a log-linear plot like
Fig.~\ref{fig:zonaexp} gives, for $\eps_1, \eps_2$, and $\eps_3$,
$$
	c \approx 3.00\times 10^{-6},\, 2.53\times 10^{-6}, \mbox{ and } 2.10\times 10^{-6},
$$
respectively. Note that the average exit time is of the order of $c^{-1}$ so that the average
time in $\Wcal^c$ is of the order of $4(10)^5$ iterates. That is, there are long periods outside
the bubbles.

Correlations between the transitions $\Wcal\to\Wcal^c$ and
$\Wcal^c\to\Wcal$  must be taken into account to be able to estimate the anomalous 
diffusion exponent $\chi$ from the trapping statistics.
To measure these, we consider two random variables: say $X$, that
denotes the length of a stay in $\Wcal$; and $Y$, that measures
the length of the next trapping segment in $\Wcal^c$.
In this way we can measure the correlation between successive stays in complementary regions.

For $\eps_1$ and $\eps_2$ we found that the correlation coefficient between $X$ and $Y$ to be
small, i.e., to be inside the confidence interval at the level of $95\%$ given by
Student's law.
However, for $\eps_3$ we initially found correlations. This anomaly has an 
easy geometrical explanation: the shape of the bubble is increasingly distorted 
(by the cubic term in $\psi(z)$)
as $\eps $ grows, recall Fig.~\ref{fig:trap3D-slice}.
The implication is that the size of the domain for $\Wcal$ in \eqref{eq:stickyregion} is too small
to properly contain the trapped segments around the bubble.
If we slightly increase the size of
this domain to
$$
	\Wcal_+ =\{(x,y,z)\;:\;|x|\leq0.04,|y|\leq0.15,|z|\leq0.1\},
$$
and an analogous form for $\Wcal_-$, then the correlation 
between successive stays is again small.
This enlargement only affects short stays in
$\Wcal$ and $\Wcal^c$ due to orbits that are located on
the periphery of the bubble. Hence, it has a minor effect on the long-time trapping
statistics shown in Fig.~\ref{fig:trap3D} and the
long-time behavior of $\sigma_T$ shown in Fig.~\ref{fig:anom3D}.

\subsection{Relating anomalous diffusion to stickiness}
\label{subsect:remarks}

Our numerical experiments suggest that the action diffusion for our map 
is anomalous, recall Tbl.~\ref{tbl:Exponents}. What is the relation
between the exponent $\chi$ of $\sigma_T$ and the exponent $b$ of the exit
time distribution?
A number of previous studies of the analogous phenomena for 2D maps imply
that
\begin{equation}\label{eq:chi-b}
	\chi=2-b/2,	
\end{equation}
see e.g., \cite{Kar83, GZR88, IHKM91, ZK93, ZK94, WS96}.
However this result does not hold for our map when $\eps$ is close to one; the final column in Tbl.~\ref{tbl:Exponents} shows the deviation of $\chi+b/2$ from the expected value of $2$.
Indeed, even the sign of the relation is not correct: as $b$ increases,
$\chi$ should decrease according to \eqref{eq:chi-b}; instead it increases. 

We believe that a major reason for this disagreement is the relatively small value of $c$
in the exponential decay of the $\Wcal^c \to \Wcal$ transitions. 
The point is that even though we have iterated each initial condition up to $10^{11}$ times, 
we may still be far from observing the ``correct" asymptotic behavior.  
Indeed, the derivation of \eqref{eq:chi-b} relies
on the $\Wcal^c\to\Wcal$ transitions being fast
compared with those for $\Wcal\to\Wcal^c$. 
When $c$ is small, orbits spend more time outside $\Wcal$. Hence, for a fixed
total number of iterates, less time is spent in $\Wcal$. Thus longer 
experiments are probably needed to faithfully compute the effect of the
$\Wcal\to\Wcal^c$ transitions on $\sigma_T$.

It would be interesting to
take into account the role of the parameter $c$ in the simple two-state transport model,
especially to compute finite time corrections to an asymptotic exponent.

\section{Conclusions}\label{sect:conclusion}

In the first part of this paper we constructed a family $f_\eps$ \eqref{eq:map3D} of two-angle,
one-action, volume-preserving maps of the cylinder $\Tset^2\times\Rset$ that
smoothly projects to the three-torus $\Tset^3$. This map has fixed point
accelerator modes that are born whenever $\eps = n$.  The phase space of
$f_0$ is foliated by horizontal, rotational invariant tori, and these
persist when $\eps$ (and $\mu)$ is small according to volume-preserving
versions of the KAM theorem. Thus our model generalizes Chirikov's standard map
to the 3D volume-preserving setting. 

The accelerator modes are created by a Hopf-one bifurcation. The local behavior
near this bifurcation is modeled by the Michelson quadratic volume-preserving
map \eqref{eq:Michelson_map}. Previous studies of this map gave necessary conditions for the
appearance of a bubble regular motion around the accelerator modes.

In the second part of the paper, we assessed the diffusive properties of the
$f_\eps$ as $\eps$ varied near the first Hopf-one bifurction at $\eps = 1$. 
We found, as expected, that if there are no
accelerator modes, the action variable exhibits normal diffusive behavior: its standard
deviation grows as $\sqrt{T}$. However when there is a
bubble of stable orbits, the action diffusion seems to be anomalous: the standard deviation
with exponent $\chi > 0.6$. Moreover, the exit time distribution for a neighborhood of the
bubble decay as a power-law $t^{-b}$ with $b \in (2, 3)$. Our
experiments suggest that the distribution for the lengths of untrapped segments is
exponential, and that stays outside and inside the bubbles are
independent.

In this paper
we provide evidence that $\Pcal_\eps(t)\sim t^{-b}$, $b\in(2,3)$, agreeing with the results
in \cite{MJM08}. This contrasts with the exponential distribution for exit times observed for the map in \cite{SZ09}. We do not know the reason for this radical difference. 

Another important question that remains is the relation between the exponents $b$ and $\chi$. 
From our computations, this differs from the relation obtained for the 2D case, 
recall \S\ref{subsect:remarks}. We hypothesize that the reason for this is that the mean exit time
from the complement of the bubbles is too long for our numerical experiments to reach their asymptotic limit.

The observed algebraic decay of the exit time distribution seems to imply that there exist 
remnants of destroyed invariant two-tori in the chaotic zone outside the KAM-bubble.
These would be analogous to the Cantori for 2D twist maps. There is no
theory, however, for the existence of these in the volume-preserving context.
If one could find these remnants, and compute the flux through them, then it 
should be possible to construct a Markov tree model, similar to that in 
\cite{MO86}, that could explain the observed stickiness of the bubble. 
To solve these problems requires a theory for the
destruction of invariant tori \cite{Mei12, FM13}. Is there an analogue of Chirikov's
overlap criterion? Are there remnant tori, and if so, what is their topology?

\section*{Acknowledgments}

NM, CS, and AV were supported by grants MTM2016-80117-P (Spain) and
2014-SGR-1145 (Catalonia). JDM was supported by NSF grant DMS-1211350 (USA).
We thank J. Timoneda for maintaining the
computing facilities of the Dynamical Systems Group of the Universitat de
Barcelona, that have been largely used in this work. This work started while
NM was visiting JDM at the University of Colorado Boulder under the
scholarship EEBB-I-15-10119 (MINECO, Spain). Fruitful discussions with R.
Easton are also acknowledged. 

\bibliographystyle{abbrv}
\bibliography{referencies}

\appendix
\newpage

\section{A choice for $\psi(z)$}\label{app:choicepsi}

Here we construct a concrete example of an odd, degree-one circle map $\psi$
that satisfies \eqref{eq:asspsi}.  This will be used in
\S\ref{sect:diffusion3D} to give numerical evidence of anomalous diffusion in
the dynamics of the map $f_\eps$ along the action variable.

\begin{figure}[ht]
\begin{center}
\begin{tabular}{cc}
	\includegraphics[width = 0.45\textwidth]{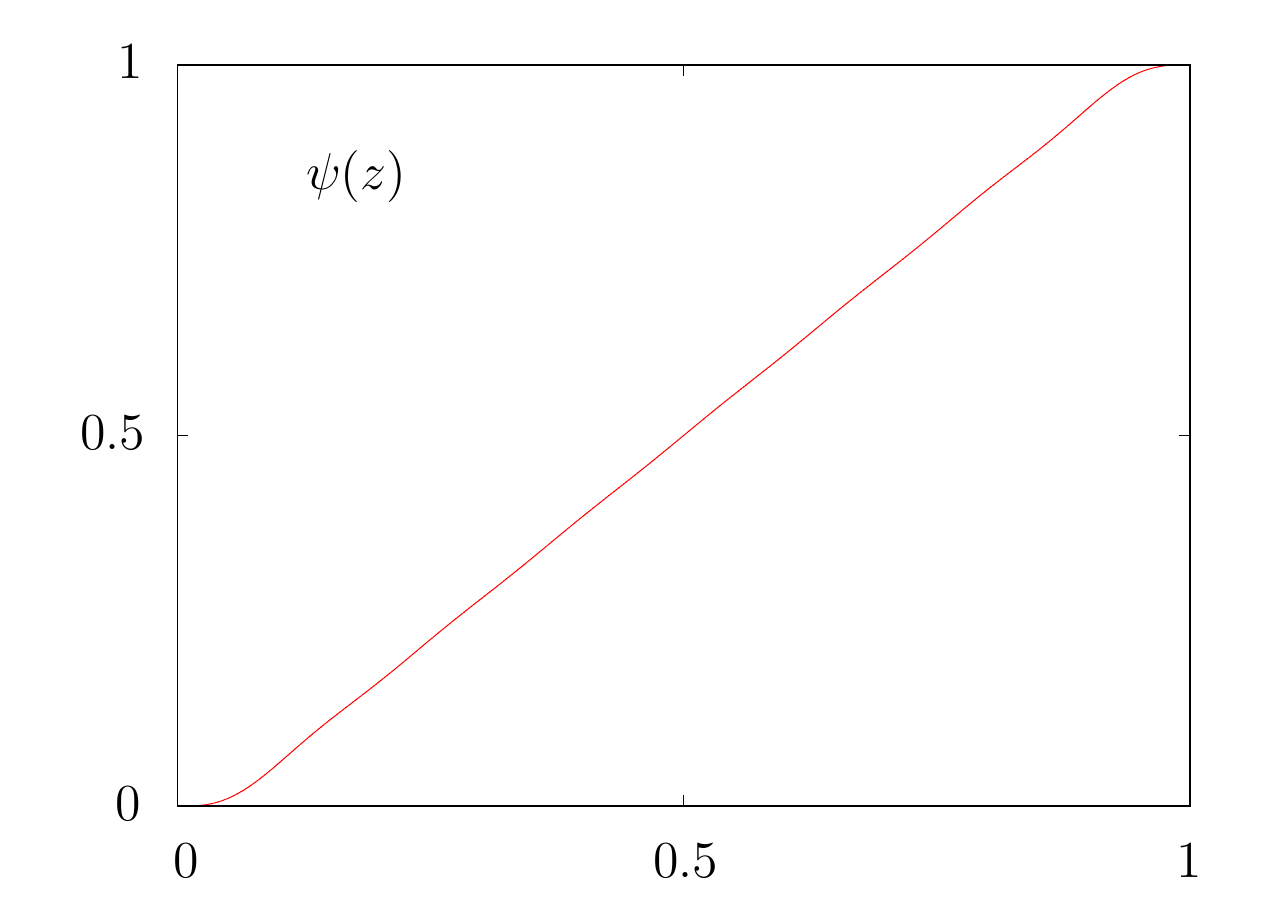}&
	\includegraphics[width = 0.45\textwidth]{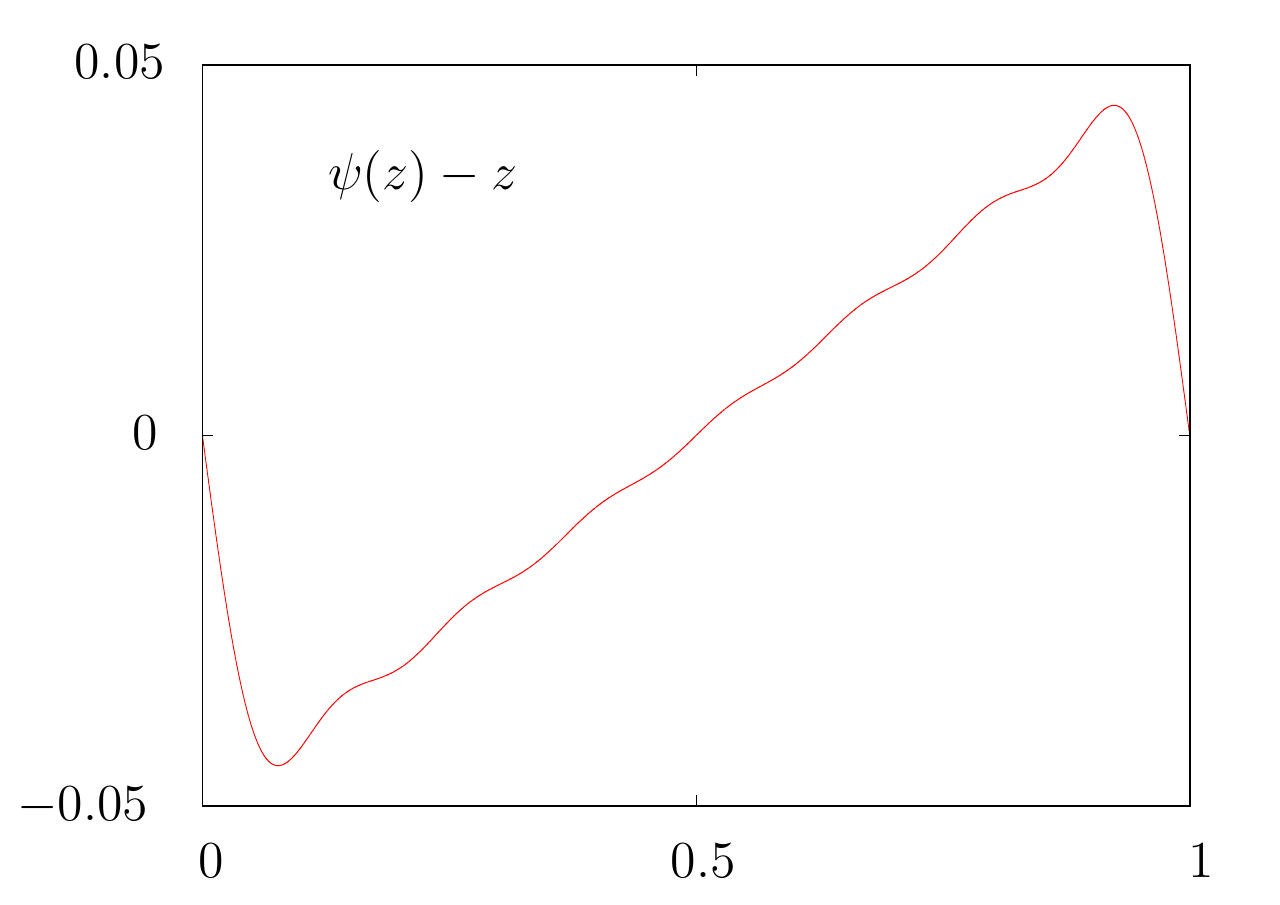}
\end{tabular}
\end{center}
\caption{Left: function $\psi(z)$ in \eqref{eq:map3D}, see \eqref{eq:psi(z)}.
Right: $\psi(z)-z$.}
\label{fig:psi(z)}
\end{figure}

First consider a function $\tilde{\psi}(z)=-z+c_3z^3$ defined on $[0,1]$. If
$c_3\geq4$ there is a unique $z_c\leq\tfrac12$ such that
$m_c=\tilde{\psi}'(z_c)$ is the slope of the straight line between
$(z_c,\tilde{\psi}(z_c))$ and $(\tfrac12,0)$. The value $z_c$ is determined as
a solution of the cubic equation,
$$
\tilde{\psi}'(z_c)(\tfrac12-z_c)+\tilde{\psi}(z_c)=0.
$$
Define the $C^1$ function 
$$
\tilde{\psi}_{\rm ext}(z)=
\left\{
\begin{array}{lcl}
\tilde{\psi}(z)   &\mbox{ if }   &z\in[0,z_c),\\
m_c(z-\tfrac12)        &\mbox{ if }   &z\in[z_c,1-z_c],\\
-\tilde{\psi}(1-z)&\mbox{ if }   &z\in(1-z_c,1].
\end{array}
\right.
$$
This is an odd function with zero average. We can consider an analytic
approximation of it via (a truncated) Fourier series, that will only contain
sine terms with coefficients $\hat{a}_k<0$. Call such an approximation
$\hat{\psi}_{\rm ext}$.  For the choice $c_3=8\pi^2$ it is
enough to take the first seven harmonics to get a fairly good approximation of
$\tilde{\psi}_{\rm ext}$. That is, we take
$$
\psi(z) = z+\lambda_c\hat{\psi}_{\rm ext}(z)
	\approx z+\lambda_c\sum_{k=1}^7\hat{a}_{k}\sin(2\pi kz),
$$
where $\lambda_c=|d\hat\psi_{\rm ext}(0)/dz|^{-1}$ is a correction factor to
make sure that $\psi'(0)=0$. For our map \eqref{eq:map3D}, this gives the form
\eqref{eq:psi(z)} with $a_i = \lambda_c \hat{a}_i$, $i = 1,\ldots,7$ being 
\begin{equation}\label{eq:aValues}
\begin{array}{ccccccc}
a_{1} &=& -0.03172255262410020, & & a_{5} &=& -0.00394622128219923,\\
a_{2} &=& -0.01500144672104500, & & a_{6} &=& -0.00257376369649251,\\
a_{3} &=& -0.00909490284466739, & & a_{7} &=& -0.00159954483407287.\\
a_{4} &=& -0.00594357151581041, & &       & &
\end{array}
\end{equation} 
In Fig.~\ref{fig:psi(z)} we can see the graph of $\psi(z)$ in $[0,1]$ (left),
and how much it differs from the identity (right).

\section{Proof of Lemma \ref{lema:FPAM}}\label{app:FPAM}

Here we prove Lemma \ref{lema:FPAM}, on the existence of fixed point
accelerator modes for the map \eqref{eq:map3D}.  Recall that the function
$\psi(z)$ is assumed to be an odd, degree-one circle map that satisfies
\eqref{eq:asspsi}.

A point $(x,y,z)$ belongs to an FPAM if $(x',y',z') =
(x+n_1,y+n_2,z+n_3)$, $n_1,n_2,n_3\in\Zset$, and $n_3\neq0$. From \eqref{eq:map3D}
this implies
\begin{eqnarray}
\mu\sin(2\pi y)+\psi(z')&=& n_1, \label{eq:comp1}\\
\nu\sin(2\pi z')        &=& n_2, \label{eq:comp2}\\
\eps\left(\cos(2\pi x)-\beta\sin(2\pi y)\right) &=& n_3. \label{eq:comp3}
\end{eqnarray}
Given the limits \eqref{eq:munuRange}, \eqref{eq:comp2} implies that $n_2=0$,
and thus either $z'=p$ or $z'=p+\tfrac12$, for some $p\in\Zset$.  
\begin{enumerate}
	\item Assume first that $z'=p\in\Zset$. Since $z'-z=n_3$, then,
$z=q=p-n_3\in\Zset$. Since $\psi(p)=p$, and $\mu$ is restricted by
\eqref{eq:munuRange}, \eqref{eq:comp1} requires that $n_1 =p$, which requires
$y = y_{\pm}$ with $y_+ =r$ or $y_- =r+\tfrac12$, for $r\in\Zset$. 

In particular, in both cases \eqref{eq:comp3} reduces to $\eps\cos(2\pi
x)=n_3\in\Zset\setminus\{0\}$. Solutions to this equation are born at $\eps =
n_3$ at $x = s$ or $x = s+\tfrac12$, being $s \in \Zset$. Hence we have FPAM
that are born when $\eps = n_3$ at the points
\begin{eqnarray*}
	P_+ =(0,0,0),& \quad &Q_+=(0,\tfrac12,0), \\
	P_- =(\tfrac12,0,0),& \quad & Q_-=(\tfrac12,\tfrac12,0),
\end{eqnarray*}
on $\Tset^3$, and all equivalent lifts of these points to $\Rset^3$.

At the Hopf-one bifurcation, the linearization $Df_\eps$ should have $1$ as
eigenvalue. This holds since at the FPAM, $\cos(2\pi y_\pm) = \pm 1$ and
$\sin(2\pi x)=0$, and the first and second traces of $Df_\eps$ are
$$
\tau   = \sigma = 3 \mp 4\beta\eps\pi^2\nu.%
$$
Finally, the second pair of multipliers is on the unit circle when $-1 < \tau =
\sigma < 3$, which gives the requirement
$$
	0 < \pm \eps\pi^2 \beta\nu < 4
$$
Thus if $\beta\nu > 0$ only the fixed points $P_{+,-}$ have the stability
property to become saddle-foci, recall Rem.~\ref{rema:rema1-3D}.

	\item If $z'=p+\tfrac12$, $p\in\Zset$, then 
since $\psi(z)-z$ is a period-one, odd function,
$\psi(z')=\psi(p+\tfrac12)=p+\psi(\tfrac12)=p+\tfrac12$.  Thus \eqref{eq:comp1}
requires that $n_1 = p+\tfrac12 +\mu\sin(2\pi y)=0$. Under the restriction
\eqref{eq:munuRange}, this implies that $n_1 \notin \Zset$. Hence, no point in
$\Tset^3$ of the form $P=(x,y,\tfrac12)$, $x,y\in\Sset^1$ can be an FPAM.  
\end{enumerate} \fin

\end{document}